\documentclass{article}

\usepackage{amssymb}
\usepackage{amsfonts}
\usepackage{epsf}
\usepackage{amsmath}

\newcommand{\vb}[1]{{\mathbf{#1}}}
\newcommand{\lb}[1]{\label{#1}}

\newcommand{\bc}{\begin{center}}
\newcommand{\ec}{\end{center}}
\newcommand{\be}{\begin{equation}}
\newcommand{\ee}{\end{equation}}
\newcommand{\bea}{\begin{eqnarray}}
\newcommand{\eea}{\end{eqnarray}}
\newcommand{\ba}[1]{\begin{array}{#1}}
\newcommand{\ea}{\end{array}}

\newcommand{\bt}[1]{\begin{table}[ht]\centering\begin{tabular}{#1}}
\newcommand{\et}[1]{\end{tabular}\caption{\small#1}\end{table}}
\newcommand{\fig}[3]{\begin{figure}[htb]\epsfxsize=80mm\bigskip\centerline{\epsfbox{#1}}\caption{\small\it #2 \label{#3}}\bigskip\end{figure}}

\begin{document}

\begin{center}

{\ \\[5mm]\Large\bf\sc\noindent  Landau-Ginzburg Chern-Simons model with\\ $U_e(1)\times U_g(1)$ Gauge Symmetry and Internal Pseudo-Photons\\[20mm]}

{\bf P. Castelo Ferreira\footnote{e-mail: \texttt{pedro.castelo.ferreira@gmail.com}\\At time of submission the author was also affiliated to: CENTRA -- Instituto Superior T\'ecnico; Av. Rovisco Pais 1; 1100-001 Lisboa -- Portugal}}\\[15mm] {\small  Grupo de F\'{\i}sica Matem\'atica -- Universidade de Lisboa\\Instituto para a Investiga\c{c}\~ao Interdisciplinar,
Av. Prof. Gama Pinto, 2; 1649-003 Lisboa -- Portugal\\[10mm]  Eng. Electrot\'ecnica -- FECN -- Universidade Lus\'ofona de Humanidades e Tecnologia\\ Campo Grande 376; 1749-024 Lisboa -- Portugal}\\[25mm]

{\bf\sc Abstract}
\end{center}
\noindent In this article it is studied, at variational level, a mathematical setup given by the Landau-Ginzburg Chern-Simons model for anyons in 2+1-dimensions within the framework of dimensional reduced $U_e(1)\times U_g(1)$ extended electromagnetism with both vector gauge fields (photons) and pseudo-vector gauge fields (pseudo-photons) such that both magnetic and electric vortexes coexist in the planar system. This model exhibits explicit planar $P$ and $T$ discrete symmetries being the Hall conductivity consistently a tensor and the Dirac quantization on the electric and magnetic coupling constants is equivalent to the quantization of magnetic flux. It is also discussed a thickening to 4-dimensions of the model with explicit 4-dimensional $P$ violation which allows either for electric and magnetic charge separation, either for the Meissner effect. Although mathematically consistent, the electromagnetic field content for this model does not coincide with the standard Hall effect being present an extra orthogonal electric and longitudinal magnetic fields.

\thispagestyle{empty}
\newpage
\tableofcontents

\newpage

\section{Introduction}

In this article it is derived a mathematical setup similar to the model for the macroscopical Fractional Hall Effect. Here we consider extended $U_e(1)\times U_g(1)$ electromagnetism containing both photon and pseudo-photon gauge fields. In the remaining of the introduction are reviewed
some theoretical results concerning the Fractional Hall Effect, its description in terms of a $U(1)$ Landau-Ginzburgh Chern-Simons model
with dynamical internal anyon and photon fields as well as are discussed the specific problems being addressed in the present work, specifically
the violation of Parity by the solutions to the equations of motion and the fractional charge values for the standard $U(1)$ models.
It is also reviewed pseudo-photon
theory in 4-dimensions and 3-dimensions. In section~\ref{sec.model} it is build a Langrangian model for the Landau-Ginzburg Chern-Simons model with an internal
dynamical pseudo-photon field and are computed and solved the respective equations of motion. Within the framework of this new model
it is also shown the existence of both electric and magnetic vortexes and derived the equivalence between Dirac's quantization condition and quantization
of magnetic flux. In section~\ref{sec.4D} it is discussed the embedding of the planar
system in the 4-dimensional manifold and estimated the statistical charge values from the perspective of the 4-dimensional manifold.
In particular it is derived a thick model for which both the electric and magnetic 3-dimensional vortexes correspond to
4-dimensional charge configurations. Are also discussed the 4-dimensional and 3-dimensional Lorentz discrete symmetries $P$, $T$ and $C$.

\subsection{Quantum Hall Effect and Fractional Statistics}
The integer Hall effect was first analyzed experimentally by Klitzing, Dorda and Pepper~\cite{IHE_exp} in 1980 and
explained theoretically by Laughlin~\cite{Laughlin_1}. For high mobility planar electron systems
at low temperature under an orthogonal (to the planar system) strong magnetic field $B$, when a longitudinal
 (along the direction of the planar system) electric field $E_i$ is applied it is induced a longitudinal transverse
(orthogonal to the applied external electric field) electric current, the Hall current.
The fractional Hall effect was unexpectedly detected in 1982 by Tsui, Stormer and Gossard~\cite{FQHE_exp}, who measured a Hall current
\be
J_H^i=\sigma_H\,\epsilon^{ij}\,E_i\ ,\ \ \sigma_H=\frac{1}{3}\,\frac{e^2}{h}=\frac{1}{3}\,\frac{e}{2\Phi_0}\ ,
\lb{J_Hall}
\ee
where $\epsilon^{ij}$ is the levi-civita symbol ($\epsilon^{xy}=-\epsilon^{yx}=1$ and $\epsilon^{xx}=\epsilon^{yy}=0$),
$e$ is the unit electric charge, $h$ the Planck constant, $\Phi_0=h/2e$ the magnetic flux quantum and $\sigma_H$ is known as the
Hall conductance. The measured value of the Hall conductance~(\ref{J_Hall}) corresponds to the fractional filling of the lower Landau
level of $\nu_{3}=1/3$, hence named fractional Hall effect. The Laughlin wave function for the fractional Hall effect~\cite{Laughlin_2}
was at that time derived phenomenologically and it is the best account for this effect as well as renders very low energies for the wave
function solutions. Considering complex coordinates for the fractional filling fraction $\nu_{m}=1/m$ ($m$ being an odd integer), up to a
normalization constant ${\mathcal N}_m$, explicitly the Laughlin wave function is
\be
\psi_{m}={\mathcal N}_m\prod_{j<k}(z_j-z_k)^m\,e^{-\frac{1}{4}\sum_l\bar{z}_lz_l}\ ,
\lb{lgwv_m}
\ee
according to the original interpretation $j,k,l$ label the positions of the several quasi-hole excitations in the system,
each with fractional electric charge $e^*=\nu_m e=e/m$.
The state $\psi_{m+1}$ is obtained by adding further quasi-holes to the system. In order to do so
it is applied the quantum creation operator $\Pi_i(z_i-z_0)$ which has the effect to pierce the Hall system by
one flux quantum at $z_0$ (for the state $m+1$ are required as many quasi-hole insertions as the ones already existing in
the system for state $m$ inserted at positions $z_l$). We note that in this construction the electrons are
considered a quantum fluid and these hole excitations are interpreted as \textit{bubbles}
in the fluid. Consistently with this interpretation they decrease the energy of the quantum state,
hence are usually termed ghosts or phantoms. Specifically the classical potential energy corresponding
to the quantum state $\psi_m$ is $m/2\sum_l \bar{z}_lz_l-2m^2\sum_{j<k}|z_j-z_k|$~\cite{Laughlin_2}.
In addition there exist also quasi-particles excitations in the system. Their quantum creation operator is
$\Pi_i(\partial/\partial z_i-z_0/a_0)$ for $a_0=\sqrt{2}\lambda_D$ being the magnetic length
of the electron fluid (or OCP plasma).

Soon it was derived that these quasi-holes and quasi-particles do not obey either a Fermi or a Bose spin-statistics relation, instead have
a fractional spin-statistics relation~\cite{frac_1,Haldane,Alperin,frac_4,frac_5}. Directly from the microscopical wave
function~(\ref{lgwv_m}), by considering the effect of adiabatically rotating quasi-particles around each-other
is obtained a Berry phase rotation~\cite{Berry} of $\Delta\gamma=2\pi\nu$. Directly
identifying this rotation with the Aharanov-Bohm phase rotation~\cite{Simon,AB} due to an auxiliary
gauge field $e^*\oint \vb{a}\cdot\vb{dl}=2\pi e^*\Phi/e\Phi_0$, we obtain the solution (for further details see the original work
of Arovas, Schrieffer and Wilczek~\cite{frac_4} and references therein)
\be
e^*\oint \vb{a}\cdot\vb{dl}=2\pi\nu\ \Rightarrow\ a^i=\frac{\nu\Phi_0}{2}\,\frac{\epsilon^{ij}(r_j-\bar{r}_j)}{(\vb{r}-\vb{\bar{r}})^2}\ .
\lb{vortex_0}
\ee
Here $\vb{\bar{r}}$ stands for the centre of the vortex,
$\Phi$ is the total flux of the vortex and it is being considered the fractional filling fraction
$\nu_m=1/m$ which correspond to cusps in the phase space~\cite{Haldane,Alperin}.
We note that this result is obtained considering that quasi-particles have spin 0 and that,
microscopically, the spin-statistics relation is directly related to conservation of angular momenta~\cite{spin_1} such that,
consistently with the original derivation of Laughlin wave function, the wavefunctions~(\ref{lgwv_m})
are eigenfunctions of angular momentum. This is the main argument to justify that only odd values of $m$ are allowed.

After these developments similar results have been obtained employing an effective Landau-Ginzburg
Chern-Simons models for the quasi-particles. The main motivations for this construction is to derive
a macroscopical model at action level that can explain the Laughlin wave function and the Hall effect
from a more fundamental variational principle. Also we recall that the interactions between fermions is usually
mediated by electromagnetic fields, as well as most systems are controllable by external fields and usually also induce
new fields due to the internal interactions of the system which can, in principle, be measured. As a simple example
we recall that when a electric charge is present its presence is known by measuring the respective electric field.
Hence a full description of any electronic system must also account for both external and induced gauge fields.

In these gauge field models anyons are commonly considered as
spin~0 bosons\cite{footnote1}
interpreted as composite fermions constituted by one electron with magnetic vortexes
attached. Similarly to the models describing superconductivity~\cite{SC_1,SC_2,SC_3}
anyons are represented by a scalar complex field $\phi$. For the Fractional Hall effect
it is further considered a Chern-Simons
term~\cite{mass_CS} for a collective gauge field (interpreted as an effective statistical gauge
field)~\cite{CS_1,CS_2,CS_3,Jain}. In these macroscopical models the fractional spin-statistics relation
for anyons is set by the value of the Chern-Simons coupling.
Purely at macroscopical model level this choice is imposed externally not being derived from the model~\cite{spin_2,spin_3,spin_4},
the Abelian Chern-Simons coupling is, generally, a free parameter. It is important to stress that in the original
derivation of Chern-Simons models from the microscopical description the value of the Chern-Simons coupling is set
either by considering the macroscopical partition function~\cite{CS_1,CS_2} (in which case the Chern-Simons term is
a Lagrange multiplier term obtained in the thermodynamical limit) or directly by relating the macroscopical
and microscopical Hamiltonians~\cite{spin_2,spin_3}. However the reverse statement is not
generally obeyed such that the macroscopical models, just by themselves, do not fix the spin-statistics relation.
Nevertheless non-trivial topologies impose further constraints in the charge spectrum as has been derived for
example in~\cite{Semenoff} (for $3+1$-dimensions). Also considering several charge insertions constraint the
allowed charge spectrum of the theory due to the braiding of Wilson lines~\cite{Polyakov}. Chern-Simons and Maxwell
Chern-Simons have been often used as a gauge description of conformal field theories\cite{CFT}, as for example
string theory~\cite{Kogan,BCK,Horava,PCT,Dbranes}. In all these frameworks the value of the Abelian Chern-Simons
coupling is fixed externally depending on the physical system that we are describing.

\subsection{Generic Filling Fractions $\nu_{n,m}=n/m$}

So far we have not discussed generic filling fractions $\nu_{n,m}=n/m$ and the respective wave functions $\psi_{n,m}$.
We recall that in the original derivation only fractions of the form $\nu_{1,m}=1/m$ were considered, which
correspond to the wave function solution~(\ref{lgwv_m}). Based in this solution, two main approaches developed by
Jain~\cite{Jain} and Haldane and Alperin~\cite{Haldane,Alperin} have been considered to describe other rational
filling fractions.

Jain considered, for a given wave function $\psi_{1/p}$ (with $\nu_{p,1}=p$), to add to each electron in the system
further $2q$ flux tubes~\cite{Jain}. Up to a normalization constant
${\mathcal N}_{p,q}$, the resulting wave function is
\be
\Psi_{1/p,2qp\pm 1}={\mathcal N}_{p,q}\Pi_{i<j}(z_i-z_j)^{2q}\psi_{1/p}={\mathcal N}_{p,q}\Pi_{i<j}(z_i-z_j)^\frac{p}{2qp\pm 1}e^{-\sum_l z_l\bar{z}_l/4}\ .
\lb{lJain}
\ee
Hence these wave functions describe a \textit{new} state with filling fraction $\nu_{p,2qp\pm 1}=p/(2qp\pm 1)$,
where the $\pm 1$ depends on the flux direction of the $2q$ vortexes added to the system. These states correspond to
excitations of charge $e^*=\nu_{p,2pq\pm 1}\,e$ and for several combinations of $p$ and $q$ all the odd denominator
fractions are obtained.

Although based in the same wave solution~(\ref{lgwv_m}), Haldane and Alperin considered a slightly
different construction, for which the quantum states correspond to excitations of both quasi-holes
and quasi-particles~\cite{Haldane,Alperin}. Their trial wave function is
\be
\Psi_s={\mathcal N}_{s}P_s\,Q_s\,e^{-\sum_i z_i\bar{z}_i/4}\ \ ,\ \ P_s=\Pi_{i<j}|z_i-z_k|^{2p_{s+1}}\ \ ,\ \ Q_s=\Pi_{i<j}|z_i-z_k|^{\frac{\alpha_s}{r_s}}
\lb{lHH}
\ee
with $p_{s+1}$ an arbitrary integer, $r_s$ an odd integer, $e_s$ the quantum excitation charge and $\alpha_s=\pm 1$
depending whether we are dealing with particle or hole excitations. The complex coordinate defined as $z=x\pm iy$ with
the sign of the imaginary part depending on the charge of the quasi-particle/quasi-hole being positive or negative.
The recursion relations for these solutions are $r_{s+1}=2p_{s+1}-\alpha_{s+1}/r_s$ and
$e_{s+1}=\alpha_{s+1}e_s/r_{s+1}$ taking the several possible specific combinations of
the integers $p_{s+1}$ and $\alpha_{s+1}=\pm 1$ assuming the initial conditions $\alpha_1=r_0=e_0=1$.
The filling fraction for these states is $\nu_{s+1}=\nu_s+\alpha_{s+1}e_{s+1}|e_{s+1}|/m_{s+1}$
with the initial condition $\mu_0=0$. The first iteration, for generic $p_{s=1}=p$, gives $r_{s=1}=2p-1$, $\nu_{s=1}=1/(2p-1)$
and $e_{s=1}=1/(2p-1)$. Hence from these recursion relations it is straight forward to conclude that, at each level $s$
of the recursion relation, corresponding to a filling fraction $\nu_s=\nu_{n,m}=n/m$, the respective charge is $e^*=e_s=\pm 1/m$.

From both constructions introduced by Jain and Haldane and Alperin, we obtain distinct hierarchies for the fractional Hall states.
The crucial difference between both the hierarchies is that Jain hierarchy is based in a vortex description of the theory while
Haldane-Alperin hierarchy is based in the quasi-particle and quasi-hole creation operators. Although both these two hierarchies are
based in the Laughlin wave function~(\ref{lgwv_m}) corresponding to projections of the excited wave-functions into the lowest Landau level
(see the original references~\cite{Jain,Haldane,Alperin}
for a more detailed discussion), they correspond to distinct interpretation of the physical
fractional charge for each quantum excitations in the Hall system, while in Jain hierarchy the fractional charges are
generic fractions of odd denominator, in the Haldane-Alperin hierarchy the fractional charge has always numerator $1$ and an odd integer
denominator.

Prior to these developments have been also considered effective theories with pure gauge Chern-Simons theories having several gauge
fields~\cite{K_CS_1,K_CS_2,K_CS_3,K_CS_4,K_CS_5} corresponding to a gauge symmetry group $U(1)^N$.
These theories are interpreted as an effective description of the Hall effect containing
only gauge degrees of freedom (the anyon field is integrated out from the theory). The several
$N$ gauge fields are interpreted as multiple collective gauge fields and physically correspond to
the superposition of several standard photon fields. Here standard means that there is no $N$
physical distinct types of charges in the theory, each of the charge of the $N$ $U(1)$ gauge groups correspond,
physically, to the usual electric charge (or more generally to fractional electric charges) such that
this construction accounts for the superposition of many electromagnetic fields.
We further note that the Jain Hierarchy and the Haldane-Alperin hierarchy are, generally, not
equivalent to each other and only within the framework of pure Chern-Simons theories was possible to show their equivalence~\cite{K_CS_3}.

For reviews in the several topics addressed in this section see~\cite{review}.

\subsection{Landau-Ginzburgh Chern-Simons Models as a Macroscopical Description of the Fractional Hall Effect\lb{sec.int.CS}}

Next we briefly resume the description of quantum Hall planar systems by effective Chern-Simons Landau-Ginzburg models. Generally, in these frameworks, are considered macroscopical $2+1$-dimensional
gauge theories containing both external fields $A_\mu$ (the fields applied externally to the planar
system) and internal gauge fields $a_\mu$ coupled to scalar particles $\phi$ describing the
anyons (usually assumed to be bosons, as already mentioned).

\subsubsection{Statistical Partition Functions and Variational Principle Formulation}

Let us start by reviewing how the statistical charge and current densities are obtained.
For a given planar Lagrangian $\mathcal{L}_0[A,a,\phi]$ describing a Landau-Ginzburgh Chern-Simons model, the
partition function is defined as a functional of the external fields $A_\mu$
\be
Z[A]=\int{\mathcal{D}}a{\mathcal{D}}\phi\,e^{-iS[A,a,\phi]}\ \ ,\ \ S[A,a,\phi]=\int dx^3{\mathcal{L}}_0[A,a,\phi],
\ee
where $S$ stands for the action and here are considered only the 3-dimensional coordinates $x^\mu$, $\mu=0,1,2$.
As usual the physical state of the system corresponds to the
minimum of the action, hence to the solutions of the equations of motion for the internal
fields, $\delta{\mathcal{L}}_0/\delta a_\mu=\delta{\mathcal{L}}_0/\delta\phi=0$.

For an external observer the controllability of the planar system is achieved by controlling the external fields $A_\mu$,
more specifically the orthogonal magnetic field (applied orthogonally to the planar system) and the longitudinal electric field
(applied along the planar directions of the Hall system)
and 
\be
\ba{rclcl}
E^i&=&F^{0i}&=&\partial_0 A_i-\partial_i A_0\ \ ,\ \ i=1,2\ ,\\[3mm]
B^\perp&=&\epsilon^{ij}F_{ij}/2&=&\partial_1A_2-\partial_2A_1\ .
\ea
\ee
We stress that in standard 2+1-dimensional (planar) electromagnetism (and respective $U(1)$ gauge theories, either Maxwell, Chern-Simons and Maxwell Chern-Simons)
these three components ($E^1$,\ $E^2$ and $B^\perp$) are the only non-null components of the electromagnetic fields. 

The measurability of the system reaction is usually achieved by measuring
the induced electromagnetic current and charge densities. Within each system the theoretical estimative for the values of these currents
are computed from the partition function for each state of the system by computing the statistical electromagnetic currents.
Specifically it is required to compute the effective partition function for each allowed physical state taking in account the equations of motion and integrating the internal degrees
of freedom
\be
\bar{Z}[A]=\int{\mathcal{D}}a{\mathcal{D}}\phi\,\Pi_\mu\delta\left(\frac{\delta{\mathcal{L}}_0}{\delta a_\mu}\right)\delta\left(\frac{\delta{\mathcal{L}}_0}{\delta \phi}\right)\,e^{-iS[A,a,\phi]}=e^{-i\bar{S}[A]}\ .
\lb{bZ}
\ee
Then taking a reference current value $\left<J^\mu_A\right>|_{A=A_{\mathrm{ref}}}$ for a known state of the system
and computing a current for a given shift of the external fields by an amount $\Delta A$ is obtained the
value for induced statistical current $\left<J^\mu_{A,\mathrm{ind}}\right>$
\be
\ba{rcl}
\left<J^\mu_A\right>&=&\displaystyle-i\frac{\delta\log \bar{Z}}{\delta A^\mu}=\frac{\delta\bar{{\mathcal{L}}}_0[A]}{\delta A_\mu}\ ,\\[3mm]
\left<J^\mu_{A,\mathrm{ind}}\right>&=&\displaystyle-i\left(\left.\frac{\delta\log \bar{Z}}{\delta A^\mu}\right|_{A_\mu=A_{\mu,\mathrm{ref}}}-\left.\frac{\delta\log \bar{Z}}{\delta A^\mu}\right|_{A_\mu=A_{\mu,\mathrm{ref}}+\Delta A_\mu}\right)\ ,
\ea
\lb{J_Z}
\ee
where the action $\bar{{\mathcal{S}}}_0[A]$ in~(\ref{bZ}) and the Lagrangian $\bar{{\mathcal{L}}}_0[A]$
are related as usual $\bar{{\mathcal{S}}}_0[A]=\int dx^3 \bar{{\mathcal{L}}}_0[A]$. 

Next we discuss a distinct construction which allows at the level of the action, both to include the effect of the statistical currents
in the planar system, and to obtain these statistical currents directly at variational level
without the explicit construction of partition functions. This is achieved by explicitly considering the inclusion of
the electromagnetic statistical currents in the Lagrangian coupling both to the internal gauge field $a$ and
external gauge field $A$
\be
{\mathcal{L}}={\mathcal{L}}_0 - (A_\mu+a_\mu)\left<J^\mu_A\right>\ .
\label{L}
\ee
These couplings are simply justified by noting that any macroscopical electromagnetic current will
couple both to the external and internal gauge fields, hence the explicit inclusion of them in the Lagrangian
is interpreted as a \textit{backreaction} effect of these statistical currents. Also we
note that it is required for consistence between the internal fields equations of motion and the current equations
such that the statistical currents are consistently given both by the variational derivation of the Lagrangian with
respect to the internal gauge field $a_\mu$ (equations of motion) and with respect to the external gauge field $A_\mu$ (current equations)
\be
\left\{\ba{rcl}\displaystyle\frac{\delta{\mathcal{L}}}{\delta A_\mu}&=&0\\[5mm]
               \displaystyle\frac{\delta{\mathcal{L}}}{\delta a_\mu}&=&0\ea\right.\ \Rightarrow\ 
\left<J^\mu_A\right>=\frac{\delta{\mathcal{L}}_0}{\delta A_\mu}=\frac{\delta{\mathcal{L}}_0}{\delta a_\mu}\ .
\ee
Hence this formulation sets a relative constraint between the equations of motion and the current equation through the statistical currents.
We further note that this construction imposes macroscopical global charge conservation in the system, the effective 3-dimensional electric
charge densities are null, from the variational derivations of the Lagrangian with respect to $A^0$ we obtain
\be
\rho_{e,\ \mathrm{eff}}=\frac{\delta{\mathcal{L}}}{\delta A_0}=\frac{\delta{\mathcal{L}}}{\delta a_0}=0\ ,
\lb{charge_conservation}
\ee 
as required for global charge conservation. We remark however that a statistical charge density is allowed
given by $\left<\rho_e^0\right>=\delta{\mathcal{L}}_0/\delta A_0=\delta{\mathcal{L}}_0/\delta a_0$. For the particular
case for which $\left<\rho_e^0\right>=0$, the system is neutral with respect to global statistical electric charge.

\subsubsection{The Chern-Simons Model Describing the Macroscopical Fractional Hall System}

Once clarified how to estimate the measurable electromagnetic currents at variational level we
will proceed to review the Chern-Simons model for Hall systems.

The specific formulation at functional level of the fractional Hall effect was first considered by
Girvin and MacDonald~\cite{CS_1} who considered the Chern-Simons action in order to implement a
Lagrange multiplier term describing the gauge field definition~(\ref{vortex_0}) as introduced in~\cite{frac_4}.
Later was derived a full $2+1$-dimensional Lagrangian ${\mathcal{L}}_0^a$ for these theories by Zhang, Hansson and Kivelsson~\cite{CS_2}
(see also the works of Read~\cite{CS_3} and Jain~\cite{Jain})
\be
\ba{rcl}
{\mathcal{L}}^a&=&{\mathcal{L}}^a_{0} - (A_\mu+a_\mu)\left<J^\mu_A\right>\ ,\\[5mm]
{\mathcal{L}}_0^a&=&{\mathcal{L}}_{\phi,a}+{\mathcal{L}}_a\ ,\\[5mm]
{\mathcal{L}}_{\phi,a}&=&\displaystyle-\phi^*\left[i\partial_0-e(A_0+a_0)\right]\phi-\frac{1}{2\bar{m}}\phi^*\left[-i\vb{\nabla}-e(\vb{A}+\vb{a})\right]^2\phi+\mu\phi^*\phi-\lambda(\phi^*\phi)^2\ ,\\[5mm]
{\mathcal{L}}_a&=&\displaystyle\frac{e^2}{4\theta_k}\epsilon^{\mu\nu\lambda}a_\mu\partial_\nu a_\lambda\ .
\ea
\lb{L_orig}
\ee
Here $A$ is the external gauge field (imposed externally to the system) and $a$ is the
internal gauge field (the dynamical gauge field). In addition to the Lagrangian terms considered in
the original derivation~\cite{CS_2} we are including the statistical current terms~(\ref{L}). Although
this construction was not originally considered we remark that it is required for consistency between
the equations of motion and the current equations.

The equations of motion for the internal fields $a_0$, $a_i$ and $\phi$ are, respectively,
\bea
\frac{\delta{\mathcal{L}}^a}{\delta a_0}=0&\Leftrightarrow&\frac{e^2}{2\theta_k}\epsilon^{0ij}\partial_ia_j+e\phi^*\phi=\left<J^0_A\right>\ ,\label{eom_a_1}\\
\frac{\delta{\mathcal{L}}^a}{\delta a_i}=0&\Leftrightarrow&-\frac{e^2}{2\theta_k}\epsilon^{0ij}\left(\partial_0a_j-\partial_ja_0\right)-\frac{e}{\bar{m}}\phi^*\left[i\partial_i+e(A_i+a_i)\right]\phi+\frac{ie}{2\bar{m}}\partial_i(\phi^*\phi)=\left<J^i_A\right>\ ,\label{eom_a_2}\\
\frac{1}{\phi}\frac{\delta{\mathcal{L}}^a}{\delta \phi^*}=0&\Leftrightarrow&-\frac{i}{\phi}\partial_0\phi+e\left(A_0+a_0\right)-\frac{1}{2\bar{m}}\frac{1}{\phi}\left[i\partial_i+e(A_i+a_i)\right]^2\phi+\mu-2\lambda\left(\phi^*\phi\right)=0\ .\label{eom_a_3}
\eea
Assuming homogeneous static solutions for the anyons and electromagnetic fields such that $\partial_0\phi=\nabla_i\phi=0$ and constant $B$ and $E^i$,
from the equation of motion for the anyon~(\ref{eom_a_3}) we obtain
\be
a_\mu=-A_\mu\ \ ,\ \ \frac{\mu}{2\lambda}=\phi^*\phi=\textrm{constant}=N\ ,
\lb{sol_static}
\ee
where $N$ represents the number of anyons in the system with $\phi=\sqrt{N}e^{i\varphi}$ and a constant phase $\varphi$,
and $\mu=2\lambda\,N$. From the Gauss' law, the equation of motion for $a_0$~(\ref{eom_a_1}), we obtain the statistical electric charge
of the planar system $\left<J^0_A\right>$. In the original derivation the Hall system is assumed to be electrical neutral,
hence this statistical charge must be null, $\left<J^0_A=0\right>$. Further, from this last equation we obtain two relevant results,
first the solution for the gauge field $a^i$ corresponds to the vortex solution~(\ref{vortex_0}),
secondly, the stationary solutions for the equations of motion~(\ref{sol_static}) require the external magnetic
field to be locked to specific values proportional to the number of anyons in the system
\be
b=-\frac{1}{2}\,\epsilon^{0ij}\partial_ia_j=\frac{\theta_k}{e}\,|\phi|^2\ \Leftrightarrow\ 
\left\{\ba{rcl}
a^i(r)&=&\displaystyle-\frac{\theta_k\epsilon^{0ij}}{\pi\,e}\int {dr'}^2\frac{r_j-r'_j}{|r-r'|^2}\phi^*\phi\ ,\\[5mm]
B&=&\displaystyle \frac{\theta_k}{e}\,N\ .
\ea\right.
\lb{ai}
\ee
As already mentioned the solution for the internal gauge field components $a^i$ was first introduced
in~\cite{frac_4} as a fictional gauge field, however here the field $a^\mu$ is interpreted as an effective statistical
(or macroscopical) physical field due to the many quasi-particle in the system instead of a fictional gauge field.
The spin-statistics relation for the model can be explicitly computed by considering anyons to be spin~0 bosons and performing an adiabatic rotation
which induces a Ahranov-Bohm phase shift~\cite{Polyakov} of $\Delta\gamma=\pi/\theta_k$
(also derivable using the rotation operators applied to the wave functions~\cite{spin_2,spin_3,spin_4}) such that the fractional
spin-statistics relation for anyons is set by fine-tuning the Chern-Simons coefficient $\theta_k$.
We recall that, as already mentioned in the introduction (see discussion after equation~(\ref{vortex_0})), at model level the Chern-Simons
coupling $\theta_k$ is a free parameter. Therefore at model level, in order to account both for the fractional statistics and the fractional Hall conductance~\cite{FQHE_exp}, $\theta_k$ is fixed to be an odd integer multiplied by $1/2$
\be
\theta_k=\frac{2k-1}{2}\ \ ,\ \ k\,\in\,{\mathbb{N}}\ .
\lb{theta_k}
\ee
As for the relation between the external applied magnetic field and the number of anyons in the system it is consistent with the experimental
step profile for the Hall conductance which is driven by the external magnetic field.

Finally, with respect to the second equation of motion~(\ref{eom_a_2}) we obtain the definition of the statistical vectorial current, the Hall current
\be
\left<J^i_A\right>=-\frac{e^2}{2\theta_k}\epsilon^{0ij}\left(\partial_0a_j-\partial_ja_0\right)=+\frac{e^2}{2\theta_k}\epsilon^{0ij}\left(\partial_0A_j-\partial_jA_0\right)\ ,
\lb{hall_orig}
\ee
such that the Hall conductance is $\sigma_H^{ij}=e^2/(2\theta_k)\epsilon^{0ij}$ and the respective Lowest Landau Level filling
fraction is $\nu_k=1/(2\theta_k)$.

The current equations (the functional derivatives with respect to the external gauge field $A_\mu$) consistently hold the same
expressions for the statistical currents being, for this model, redundant. Also we note that we are considering natural units for which $h=1$,
hence there are no explicit factors of $\pi$ in the equations derived here~\cite{CS_2}.

The solutions~(\ref{sol_static}),~(\ref{ai}) and~(\ref{hall_orig}) for the equations of motion are valid for
generic configurations of the external fields. In particular it is relevant to describe the following three particular
cases corresponding to a field setup sequence for measurement of the hall conductance:
\begin{itemize}
\item[\ \ I.] \textbf{no external fields applied},\\
$E^i=B=0$, we obtain the trivial solutions without any anyons in the system and no Hall current present, hence from the
gauge field solutions $a_\mu=A_\mu=0$ we obtain
\be
N=0\ \ \ ,\ \ \ \left<J^i_A\right>=0\ ;
\lb{CS_no_fields}
\ee
\item[\ II.] \textbf{only an orthogonal magnetic field is applied},\\
$E^i=0$ and $B=\epsilon^{ij}\partial_iA_j/2\neq 0$, the external magnetic field creates anyons in the Hall system and no Hall current is present,
hence from the gauge field solution $a_0=A_0=0\ ,\ a_i=-A_i$ we obtain
\be
N=\frac{2eB}{2k-1}\ \ \ ,\ \ \ \left<J^i_A\right>=0;
\lb{CS_B_field}
\ee
\item[III.] \textbf{both an external orthogonal magnetic field and a longitudinal electric field are applied},\\ $E^i=\partial_0 A_i-\partial_i A_0\neq 0$ and $B=\epsilon^{ij}\partial_iA_j/2\neq 0$, the external electric field induces an Hall current, hence from the gauge field solution $a_\mu=-A_\mu$ we obtain
\be
N=\frac{2eB}{2k-1}\ \ \ ,\ \ \ \left<J^i_A\right>=\sigma_H^{ij}\,E_j\ ,
\lb{CS_E_B_fields}
\ee
with the Hall conductance given by $\sigma_H^{ij}=e^2/(2k-1)\epsilon^{ij}$ from~(\ref{theta_k}) and~(\ref{hall_orig}).
\end{itemize}

The remaining of the fractional conductance multiples of these fractions are
justified in this model by the existence of finite energy vortex solutions~\cite{frac_4,frac_5,Jain,Abrikosov,NO,Nambu}
\be
\phi=e^{\pm i\varphi}\ \ ,\ \ \ \ a^i=\pm\epsilon^{ij}\,\frac{r_j}{e\,|\vb{r}|}\ ,
\lb{vortex_CSa}
\ee
here expressed in Cartesian coordinates centred at the origin such that the several Hall conductance
(or equivalently the filling fractions) can be obtained by adding such vortex solutions to the system
and grouped in a hierarchy by considering the inverse Chern-Simons couplings to be
\be
\theta_{p,2k-1}=\frac{\beta}{\alpha}=\frac{2k-1}{p}\ \ ,\ \ \ p\neq 0\in{\mathbb{N}}\ ,\ \ \ k\in{\mathbb{N}}\ .
\lb{tt_kp}
\ee
where for convenience we have introduced the two constants $\alpha$ and $\beta$ obeying the following
ratio equality $\frac{2k-1}{p}=\beta/\alpha$. Strictly with respect to the fractional values of the Hall
conductance $\sigma_H=e/(2\Phi_0)\,\alpha/\beta$
(or equivalently the Landau filling factors $\nu_{p,k}=\alpha/\beta=p/(2k-1)$), the model can reproduce
either the Jain hierarchy~\cite{Jain} or the Haldane-Alperin hierarchy~\cite{Haldane,Alperin} already
discussed in the previous section. However the charge of the quantum excitations will be distinct for each hierarchy
as already notice. Let us briefly resume both original hierarchies and its possible relation with the
Chern-Simons model:
\begin{itemize}
\item[]\textit{Jain Hierarchy} -- Jain~\cite{Jain} considered that the anyons have
unit electric charge $e^*=e$ such that $\alpha=1$ and the integer filling fraction $\nu_{1,1/p}=p$ corresponds
for each $p$ electrons to have, in average, one unit of magnetic flux~\cite{footnote2,frac_4,Laughlin_3,Jain} such that $|\beta|=1/p$.
By considering further $2q$ flux tubes attached to each electron (an even integer is necessary due to the
induced phase change~\cite{Jain,frac_4}) we obtain the full Jain Hierarchy of fractional Hall conductance
and respective filling fraction $\nu_{1,(qp\pm 1)/p}=|\beta|=p/(2qp\pm 1)$, hence in equation~(\ref{tt_kp}) we have that
$k=qp$.
\item[]\textit{Haldane-Alperin Hierarchy} -- Haldane~\cite{Haldane} and Alperin~\cite{Alperin} considered that the anyons have
a fractional electric charge $e^*=\alpha e$, with $\alpha=1/(2k-1)$ coinciding with the numerator of the filling
fraction given as the inverse of the Chern-Simons coupling coefficient~(\ref{tt_kp}). Hence
with this interpretation we have, generally, that for a given filling fraction $\nu_{p,2k-1}=p/(2k-1)$ we always obtain a fractional charge $e^*=1/(2k-1)$.
\end{itemize}

Hence, the conclusion is that although we obtain the same allowed Hall conductance and filling fractions
on both hierarchies the physical interpretation for each of them is distinct. In particular the
interpretation in terms of the measurable fractional electric charge being $e^*=\alpha\,e$. While
in Jain Hierarchy we always have $e^*=e$, in Haldane-Alperin Hierarchy we have $e^*=e/(2k-1)$.
We also remark that the Chern-Simons Landau-Ginzburg model,
as described in this section, is only compatible with the Jain Hierarchy. The
electric charge of anyons is fixed to be $e^*=e$ which is explicit at the Lagrangian level by
the electric coupling constant being $e$ in the terms containing products of the gauge fields with the anyon field $\phi$.

\subsection{Problems Being Addressed}

Given the results presented so far there are two main problems concerning the Chern-Simons model just
discussed that we will address and tentatively solve employing a distinct Landau-Ginzburg Chern-Simons model:
\begin{enumerate}
\item The 3-dimensional discrete symmetries parity $P$ and time-inversion $T$ are explicitly violated both
at level of the action and of the electromagnetic equations of motion due to the relation between scalar (or vector)
quantities and pseudo-scalar (or pseudo-vector) quantities. In particular the
internal field solution for the internal gauge field $a^i$ (a vector) is given in terms of a pseudo-vector quantity
as expressed in equations~(\ref{vortex_0}) and~(\ref{ai}), as well as the well-known electric Hall current $J_H^i$~(\ref{J_Hall}), a vector current,
in the presence of an external electric field $E_i$, is given by a pseudo-vector quantity, $\sigma^{ij}_H\,E_i$~(\ref{hall_orig}),
$E^i$ transforms lika a vector and $\epsilon^{ij}$ as pseudo-tensor, hence $\sigma^{ij}_H$ transforms as a pseudo-tensor and $\sigma^{ij}_H\,E_i$ as
a pseudo-vector.
\item There is experimental evidence that, for a given Hall conductance $\sigma_H=e/(2\Phi_0)\,p/(2k-1)$, the corresponding
fractional charges are $e^*=e/(2k-1)$~\cite{exp_charge}. Therefore interpreting this result in terms of
anyons we would expect that the number of magnetic vortexes (number of elementary quantum flux) per electron is $1/p$
as in Haldane-Alperin Hierarchy which does not coincide with the standard interpretation for the
the Chern-Simons Landau-Ginzburg model just discussed.
\item A 4-dimensional description of the electromagnetic statistical charges and fields is desirable in order to evaluate whether
the Hall system configurations correspond to either measurable charge configurations or induced electromagnetic fields orthogonal to the
system, in particular whether it is possible to describe physically measurable fractional electric and magnetic charges of anyons.
For Maxwell or Maxwell Chern-Simons it is not possible to describe either orthogonal electric fields either orthogonal electric flux
which are required to achieve such construction.
\end{enumerate}

In the remaining of this work we will consistently solve these problems however, although motivated by the fractional Hall effect, the resulting model does not describe the standard Hall effect, instead there will exist extra electromagnetic fields. Let us proceed and briefly justify why the above statements constitute a problem and the relevance of trying to solve them.

Concerning the first point it is widely accepted that strong external fields applied to planar systems violate 4-dimensional parity.
This violation is a direct consequence of the spatial system configuration, in very simple terms the system is not a 4-dimensional system,
instead it is an approximately planar 3-dimensional system. However, with the exception of chiral phases, where this symmetry is explicitly broken due to spin polarization effects, one may expect that in an effective (meaning macroscopical) planar gauge theory parity symmetry should
be preserved. We remark that electromagnetic interactions are parity $P$, time-inversion $T$ and charge conjugation $C$ invariant, both at classical
level (Maxwell equations) and at quantum field theory level (QED). Moreover the planar 3-dimensional semi-classical configurations obtained from
the effective Landau-Ginzburg models are, macroscopically homogeneous and isotropic, hence from these arguments it is also expected that the
discrete symmetries are preserved in the planar system. In addition we note that the internal gauge field is commonly interpreted as being an auxiliary field instead
of a \textit{true} physical field and it is not directly measured experimentally in this physical framework.
However the electric current (the Hall current) is directly measured and is physically meaningful transforming as a vector quantity under the discrete symmetries.
Hence equation~(\ref{hall_orig}), although qualitatively correct, from a more fundamental point of view is inconsistent:
it is relating a physical vectorial current
with a pseudo-vector quantity, the Hall conductance tensor $\sigma_H^{ij}$ should transform
as a tensor instead of transforming as a pseudo-tensor. The most straight forward solution is to consider the Chern-Simons coupling $\theta_k$ to
be a pseudo-scalar (see for instance~\cite{Sing_2} for a similar construction), in the following we give another
solution to this problem by considering the internal gauge field to be a pseudo-vector which naturally
arises in pseudo-photon theory~\cite{CF,Sing_1,action,CCN,pseudo,planar}.

Concerning the second point we note that it is today an experimental fact that in fractional Hall systems anyons have
fractional charge as has been verified by several independent groups~\cite{exp_charge}. At model level it is
enough to consider the electric coupling constant to be $e^*$ instead of the standard electron charge $e$. However the
\textit{physical} existence of measurable fractional electric charges $e^*=\alpha e$ for each different
measurable Hall conductance $\sigma_H=e/(2\Phi_0)\,\alpha/\beta$, independently of $\beta$, seems to imply
that there exist both electric vortexes accounting for the value of $\alpha$ and magnetic vortexes
accounting for the value of $\beta$. Relating this last statement with the first point and noting that the magnetic field
is a pseudo-vector quantity and magnetic flux is a pseudo-scalar quantity we expect, consistently, the number of magnetic
vortexes per electron to be given by a pseudo-scalar quantity such that the planar discrete symmetries are preserved. In
the following we use the notation $\hat{\beta}$ to differentiate between the pseudo-scalar quantity and scalar quantity $\beta$.

Concerning the third point we note that considering both electric and magnetic vortexes in the system, besides being consistent with
the formulation of the Fractional Quantum Hall Effect in terms of fractional charged anyons, allows for a formulation of these systems
in a variational framework of a gauge field theory~\cite{Schwarz,Witten} compatible with the 4-dimensional Maxwell equations~\cite{action},
hence from more fundamental principles. In the following we are mostly working at semi-classical level with statistical average fields
and we will not derive a quantum formulation for the system, instead the statistical electric and magnetic vortex fluxes will
be encoded in the parameters of the model $\alpha$ and $\hat{\beta}$, however we note that these fluxes can be described as quantum excitations~\cite{mpl}. Specifically the existence of both kind of vortexes is consistent in the framework of dimensional
reduced $U_e(1)\times U_g(1)$ extended electromagnetism which we shortly review in the next section.

\subsection{Pseudo-Photon Gauge Theory}

Extended $U_e(1)\times U_g(1)$ electromagnetism was originally motivated by the work of Cabibbo and Ferrari~\cite{CF}
and the possibility of the existence of magnetic monopoles~\cite{Sing_1,action,CCN} such that are considered
a gauge sector corresponding to electric interactions and one gauge sector corresponding to the magnetic interactions.
This theory is also justified at variational level (meaning at action level) in the presence of non-regular external
electromagnetic fields~\cite{pseudo}. We note that in these conceptual systems the Maxwell equations including magnetic
4-currents, in particular the violation of the Bianchi identities, cannot be described at variational level by theories with only the standard
gauge field (photon). In addition we stress that although this theory has an $U(1)^2$ gauge group
is not equivalent to the $U(1)^N$ (with $N=2$) description of the superposition of $N$ electromagnetic distinct collective effects~\cite{K_CS_1,K_CS_2,K_CS_3,K_CS_4,K_CS_5}. In extended $U_e(1)\times U_g(1)$ electromagnetism we have two
distinct types of group charges, the electric charge $e$ and the magnetic charge $g$ and the respective gauge fields have
distinct transformation properties under the discrete symmetries, specifically it is considered a vector gauge field $A$ (the photon, corresponding
to the gauge group $U_e(1)$) and a pseudo-vector gauge field $C$ (the pseudo-photon, corresponding to $U_g(1)$).

The derivation of the four-dimensional action for $U_e(1)\times U_g(1)$ extended electromagnetism
was carried in~\cite{action} based in the Maxwell equations and considering only two assumptions:
\begin{itemize}
\item[] -- Electromagnetic interactions are $P$, $T$ and $C$ invariant;
\item[] -- It only exist one electric and one magnetic physical fields.
\end{itemize}
There is experimental evidence for both these statements which are inherent to the phenomenologically
derived Maxwell equations. In addition to the standard gauge field $A$ (the photon, corresponding to the electric
interactions and with gauge group $U_e(1)$) we have an extra gauge field $C$ that must transform under the
discrete symmetries $P$ and $T$ as a pseudo-vector field (the pseudo-photon, corresponding to the magnetic
interactions and with gauge group $U_g(1)$). The transformation properties for the psudo-photon field $C$
are obtained directly from the Maxwell equations~\cite{Sing_1}, either by noting that magnetic 4-currents are
pseudo-vector 4-currents (the magnetic field is a pseudo-vector field) or equivalently by noting that the charges of each gauge groups
are directly related to the topological charges of the other gauge group. In simpler terms the minimal
coupling to magnetic 4-currents in the Lagrangian is ${\mathcal{L}}_{J_g,\mathrm{4d}}=g\,C_I J_g^I$ (with $I=0,1,2,3$), since
$J_g^I$ is a pseudo-vector 4-current, to ensure $P$ and $T$ invariance of the action, the gauge field $C_I$ must
transform as a pseudo-vector. Given the above assumptions the lower order kinetic terms presented in the Lagrangian are~\cite{action}
\be
{\mathcal{L}}_{\mathrm{4d}}=-\frac{1}{4}F_{IJ}F^{IJ}+\frac{1}{4}G_{IJ}G^{IJ}+\frac{1}{4}\epsilon^{IJKL}F_{IJ}G_{KL}=-2\left(E_{\mathrm{4d}}^2-B_{\mathrm{4d}}^2\right)\ ,
\label{L_C_4D}
\ee
up to a sign choice of the topological Hopf term (the last term in the above equation) and with the electric and magnetic fields defined as
\be
E_{\mathrm{4d}}^i=F^{0i}+\frac{1}{2}\epsilon^{0ijk}G_{jk}\ \ ,\ \ B_{\mathrm{4d}}^i=G^{0i}-\frac{1}{2}\epsilon^{0ijk}F_{jk}\ ,\ i=1,2,3\ .
\ee

In planar 3-dimensional systems this theory allows a description of electromagnetism in terms of the full vectorial
electric and magnetic fields~\cite{planar}. Assuming approximately constant values of the fields across the orthogonal
direction to the planar system the bare gauge action for planar ($2+1$-dimensional)
extended $U_e(1)\times U_g(1)$ electromagnetism containing both the standard external gauge field $A$ (photon)
and an internal gauge field $C$ (pseudo-photon) is~\cite{Witten,Schwarz,planar,Horava}
\be
{\mathcal{L}}_C=-\frac{\delta_\perp}{4}F_{\mu\nu}F^{\mu\nu}+\frac{\delta_\perp}{4}G_{\mu\nu}G^{\mu\nu}+k\,\epsilon^{\mu\nu\lambda}A_\mu\partial_\nu C_\lambda\ .
\lb{S_AC}
\ee
$\delta_\perp$ stands for the thickness of the system along the orthogonal direction and the 3-dimensional indexes stand for $\mu=0,i$ with the roman
indexes running now only over the coordinates in the planar system $i=1,2$. The value of the Chern-Simons coefficient $k$ depends on the type of manifolds considered and the embedding of the 3-dimensional manifold into the 4-dimensional manifold, in particular whether only one
boundary or two boundary systems are considered. Specifically it can take the values\\
\begin{tabular}{rcll}
$k$&$=$&$\displaystyle\pm\frac{1}{2}\ :$&for 4D manifolds with only one 3D boundary and spin-structure~\cite{Witten,planar},\\[5mm]
$k$&$=$&$\pm 1\ :$&for 4D manifolds with only one 3D boundary without spin-structure or\\[5mm]
&&&for two boundary systems with spin-structure or\\[5mm]
&&&for antisymmetric orbifold planes with spin-structure ,\\[5mm]
$k$&$=$&$\pm 2\ :$&for two boundary systems without spin-structure or\\[5mm]
&&&for antisymmetric orbifold planes without spin-structure~\cite{Horava,PCT},\\[5mm]
$k$&$=$&$0\ :$&for symmetric orbifolds planes~\cite{Horava,Witten,planar,PCT},
\end{tabular}\\[5mm]
where the $\pm$ sign depends on the orientation of the 3-dimensional manifold. Except otherwise stated, in the following we are considering $k=1$
corresponding to a 3-dimensional manifold without spin-structure embedded into a 4-dimensional manifold, hence a single boundary system. In geometrical terms
we are considering a single plane embedded into an unbounded 4-dimensional manifold such that the second boundary may be considered
at spatial infinity for which the fields vanish (this is a standard assumption in electromagnetism and electrodynamics).

The electromagnetic field definitions are~\cite{planar}
\be
\ba{rclcrcl}
\tilde{E}^\perp&=&\displaystyle\frac{1}{2}\,\epsilon^{ij}\partial_iC_j&, &B^\perp&=&\displaystyle\frac{1}{2}\,\epsilon^{ij}\partial_iA_j\ ,\\[5mm]
E^i&=&F^{0i}=\partial^iA^0-\partial^0A^i&, &\tilde{B}^i&=&G^{0i}=\partial^iC^0-\partial^0C^i.
\ea
\lb{field_defs}
\ee
Here the tilde indicates that the respective physical fields are defined in terms of the pseudo-photon gauge field $C$. $\tilde{E}^\perp$ and $B^\perp$ stand for
the components of the electromagnetic fields orthogonal to the planar system and $E^i$ and $\tilde{B}^i$ to the components of the electromagnetic
field along the planar system. In the remaining of this work we drop the index $\perp$ such that $B=B^\perp$ and $\tilde{E}=\tilde{E}^\perp$.

We stress again that in planar systems the standard $U(1)$ Maxwell theory, Chern-Simons theory or Maxwell Chern-Simons
theory only describe the orthogonal magnetic field $B=B^\perp=\epsilon^{\perp jk}\partial_j A_k/2$ and the longitudinal
electric field components $E^i=F^{0i}$. A description of all the field components as given in~(\ref{field_defs})
is only possible in extended electromagnetism theories such as pseudo-photon theory.

Also an interesting and relevant result within the framework of dimensionally reduced pseudo-photon theory is that the
Chern-Simons coupling in~(\ref{S_AC}) is fixed from the higher dimensional theory such that, when considering
systems containing anyons, the spin-statistics relation cannot be set by fine-tuning this coupling.
Instead it should be set by topological arguments either at macroscopical level by considering vortex (3D) or flux tubes (4D) configurations
or at microscopical level by considering the braiding of particle trajectories (Wilson lines)~\cite{Semenoff}.
In the following we are mostly considering vortex configurations which can be lifted either to flux tubes or electromagnetic fields
in the 4-dimentional manifold.

\section{Landau-Ginzburg Chern-Simons Model with Pseudo-Photons\lb{sec.model}}

In this section we will develop a Landau-Ginzburg Chern-Simons model containing a
pseudo-photon field. Let us note that in the framework of $U_e(1)\times U_g(1)$ extended electromagnetism,
a particle carrying both electric flux and magnetic flux must couple both to the $A$ field
(through its electric flux) and to the $C$ field (through its magnetic flux). This is the case for anyons.
We also remark that 3-dimensional charge does not distinguish between 4-dimensional configurations corresponding
to orthogonal electromagnetic fields and charges, only in 4-dimensions, given the relative directions of the flux
at each side of the planar system can properly distinguish whether it corresponds to a field or charge configuration.
We will discuss this topic in detail in the next section, for now let us note that when external and induced orthogonal
fields are present a non-null statistical 3-dimensional charge (either electric or magnetic) may be obtained in the planar system.

From the standard macroscopical description of the Hall effect by Landau-Ginzburg Chern-Simons models discussed
in the introduction (section~\ref{sec.int.CS}) we will assume that the Hall conductance depends both on the average number of magnetic and electric
unit flux for each anyon in system. One possible interpretation for this assumption is that fractional electric charge is due to electric vortexes
vortexes in the system being the usual electron charge screened by some mechanism. Here we adopt this interpretation and, at model
level, we implement this screening by considering a simplified framework for which the internal gauge field $a$ is explicitly absent
from the model affecting only the magnetic flux carried by each anyon.
In the following we further assume macroscopical magnetic charge neutrality allowing non-null macroscopical electric charge.

To explicitly build a Lagrangian we are taking the following assumptions with respect to the field content and couplings of the model:
\begin{enumerate}
\item $\phi$ is a complex field representing a many particle state of anyons, i.e. composite electrons carrying each an average
electric flux $e\alpha$ (a scalar quantity)
and an average magnetic flux $g\hat{\beta}$ (a pseudo-scalar quantity). Here $e$ is the electric coupling constant, the charge of the electron,
and $g$ is the coupling constant corresponding to the pseudo-photon field $C$ which coincides with the unit of magnetic charge~\cite{action} according
to the original theory of Dirac monopoles~\cite{Dirac}. As for $\alpha$ and $\hat{\beta}$ represents the average number of electric and magnetic
unit fluxes carried by each anyon in the system;
\item $\phi^*\phi$ is real and transforms as a scalar under the discrete symmetries $P$ and $T$;
\item to lower order, the standard external photon field $A$ couples to the internal pseudo-photon field $C$ through the
Chern-Simons terms and to the electric flux of anyons, hence coupling to the anyon field $\phi$ through the effective
scalar coupling constant (non-dynamical and space-time independent) $e\alpha$;
\item to lower order, the internal pseudo-photon field $C$ couples to the standard external photon field $A$ through the Chern-Simons terms and to the
magnetic flux of anyons, hence coupling to the anyon field $\phi$ through the effective pseudo-scalar coupling constant (non-dynamical and space-time
independent) $g\hat{\beta}$. This coupling constant, being a pseudo-scalar,
ensures that the Lagrangian is $P$ and $T$ invariant~\cite{Sing_2};
\item as for the internal standard photon field $a$ is not considered as a dynamical variable and is explicitly excluded from the model.
Implicitly it is responsible for the magnetic flux $e\alpha$ attached to each anyon;
\item at model level we consider both electric and magnetic external currents $\left<J_e^\mu\right>$ and $\left<J_g^\mu\right>$ representing the
statistical currents as introduced in equation~(\ref{L}) in section~\ref{sec.int.CS}. Although macroscopical flux conservation is ensured, non-neutral
systems are allowed such that the non-null macroscopical statistical charge densities are interpreted as induced fluxes in the planar system.
\end{enumerate}

Given these assumptions we next proceed to write the Lagrangian for the model and derive the respective equations of motion and
current equations. We will also discuss the relation between the Hall conductance obtained from this model and Dirac's
quantization condition and explicitly show that the assumption of $\alpha$ and $\hat{\beta}$ being the average number of electric
and magnetic unit fluxes carried by each anyon in the system is corroborated by the solutions of the equations of motion. We briefly
discuss several allowed configurations for the model which are solutions of the equations of motion having distinct values for the
macroscopical statistical electric and magnetic charge densities.

\subsection{The Action and Equations of Motion for the Model\label{sec.electric}}

Given the above assumptions we consider the Lagrangian ${\mathcal{L}}^C$
\be
\ba{rcl}
{\mathcal{L}}^C&=&\displaystyle{\mathcal{L}}_0^C-A_\mu\,\left<J_e^\mu\right>-C_\mu \left<J^\mu_g\right>\ ,\\[5mm]
{\mathcal{L}}_0^C&=&{\mathcal{L}}_{\phi,C}+{\mathcal{L}}_{C}\\[5mm]
{\mathcal{L}}_{\phi,C}&=&\displaystyle-\phi^*\left[i\partial_0-e\,\alpha A_0-g\hat{\beta} C_0\right]\phi\\[5mm]
                      & &\displaystyle-\frac{1}{2\bar{m}}\phi^*\left[-i\vb{\nabla}-e\alpha\vb{A}-g\hat{\beta}\vb{C}\right]^2\phi\\[5mm]
                      & &+\mu\phi^*\phi-\lambda(\phi^*\phi)^2\ .
\ea
\lb{L_C_phi}
\ee
Here $\bar{m}$ is the effective renormalized mass of the anyon and ${\mathcal{L}}_C$ is given
in equation~(\ref{S_AC}) corresponding to the bare pseudo-photon theory action.

The usual partition function for this model is obtained by a functional integral over the internal degrees of freedom, i.e. over the fields $\phi$ and $C$,
\be
Z_C[A_0,A_i]=\int{\mathcal{D}}\phi{\mathcal{D}}C\,e^{-i\int dx^3 {\mathcal{L}}^C_0}\ .
\ee
In the following we will compute the statistical currents directly from the Lagrangian from the functional derivatives of ${\mathcal{L}}$
with respect to the external gauge fields components $A_\mu$ as discussed in the introduction (see equation~(\ref{L}) and discussion thereafter).
We will particularize the solutions of the equations of motion to stationary and uniform electromagnetic field configurations such that
the contribution of the kinetic terms (the Maxwell terms $F_{\mu\nu}F^{\mu\nu}$ and $G_{\mu\nu}G^{\mu\nu}$) are null, as well as
$\partial_0\phi=\partial_i\phi=0$. Hence in the following discussions, we omit the contribution of the Maxwell terms to the equations of motion.

For a given constant external electric field $E^i=\partial_iA_0$ and magnetic field $B=\epsilon^{ij}\partial_iA_j/2$,
expressed in terms of the gauge field $A_\mu$, the Equations of Motion for the internal fields are 
\bea
\frac{\delta{\mathcal{L}}}{\delta C_0}=0&\Leftrightarrow&\epsilon^{0ij}\partial_i A_j+g\hat{\beta}\phi^*\phi-\left<J^0_g\right>=0\ ,
\lb{EOM_C_0}\\[5mm]
\frac{\delta{\mathcal{L}}}{\delta C_i}=0&\Leftrightarrow&-\epsilon^{0ij}\left(\partial_0 A_j-\partial_jA_0\right)-\frac{g\hat{\beta}}{\bar{m}}\left(e\alpha\, A_i+g\hat{\beta}\,C_i\right)\phi^*\phi-\left<J^i_g\right>=0\ ,
\lb{EOM_C_i}\\[5mm]
\frac{\delta{\mathcal{L}}}{\delta \phi}=0&\Leftrightarrow&\left(e\alpha\,A_0+g\hat{\beta}\,C_0\right)-\frac{1}{2\bar{m}}\left(e\alpha\, A_i+g\hat{\beta}\,C_i\right)^2+\mu-2\lambda\,\phi^*\phi=0\ .
\lb{EOM_phi}
\eea
The negative sign in the first term in the left hand side of equations~(\ref{EOM_C_i})
is due to the swapping of indexes in the antisymmetric tensor, $\epsilon^{i0j}=-\epsilon^{0ij}$.
As for the statistical electric charges and vectorial currents are obtained from the functional derivatives of the
Lagrangian with respect to $A_\mu$
\bea
\frac{\delta{\mathcal{L}}}{\delta A_0}=0&\Leftrightarrow&\left<J^0_e\right>\,=\,\epsilon^{0ij}\partial_i C_j+e\alpha\phi^*\phi\ ,
\lb{EOM_A_0}\\[5mm]
\frac{\delta{\mathcal{L}}}{\delta A_i}=0&\Leftrightarrow&\left<J^i_e\right>=-\epsilon^{0ij}\left(\partial_0 C_j-\partial_jC_0\right)-\frac{e\alpha}{\bar{m}}\left(e\alpha\, A_i+g\hat{\beta}\,C_i\right)\phi^*\phi=0\ .
\lb{EOM_A_i}
\eea

Assuming static and homogeneous solutions for the electromagnetic fields (specifically $\tilde{E}$ and $\tilde{B}^i$)
we obtain from the equation of motion~(\ref{EOM_phi})
\be
C_\mu=-\frac{e\alpha}{g\hat{\beta}}\,A_\mu\ \ ,\ \ \ \ \phi=\sqrt{N}e^{i\varphi}\ ,\ \ \ \ \mu=2\lambda N\ .
\lb{C}
\ee
For these solutions the internal and external components of the electromagnetic fields are related through the
coupling constants $e\alpha$ and $g\hat{\beta}$
\be
\ba{lrcl}
\mathrm{orthogonal\ components\ :}&g\hat{\beta}\tilde{E}&=&-e\alpha\,B\ ,\\[5mm]
\mathrm{longitudinal\ components\ :}&g\hat{\beta}\tilde{B}^i&=&-e\alpha\,E^i\ .
\ea
\lb{sol_E_B}
\ee
We recall that for pseudo-photon theory all components of the electromagnetic fields exist in planar systems~\cite{planar}
as defined in equation~(\ref{field_defs}). For the model describing planar systems we are presently discussing the external fields
applied to the system are the orthogonal magnetic field $B$ and the longitudinal electric field $E^i$ and the internal induced fields
are the orthogonal electric field $\tilde{E}$ and the longitudinal magnetic field $\tilde{B}^i$.
 
Given the above solutions~(\ref{C}), considering both the magnetic current obtained from the equation of motion~(\ref{EOM_C_i}) and the statistical
electric current (the Hall current) from equation~(\ref{EOM_A_i}) we obtain the following relation for both currents $\left<J_e^i\right>$ and $\left<J_g^i\right>$
\be
\left\{\ba{rcl}\left<J_g^i\right>&=&\displaystyle -\epsilon^{ij}E_j\\[3mm]
\left<J_e^i\right>&=&\displaystyle +\frac{e\alpha}{g\hat{\beta}}\,\epsilon^{ij}E_j\ea\right.\
 \Leftrightarrow\ \left<J_g^i\right>\,=\,-\frac{g\hat{\beta}}{e\alpha}\,\left<J_e^i\right>
\ ,
\lb{Je_i_Jg_i}
\ee 
This result is consistent with the interpretation that the anions are composite particles carrying an average electric flux with equivalent electric
charge $e\alpha$ and an average magnetic flux with equivalent magnetic charge $g\hat{\beta}$. In particular imply that the currents can be re-expressed
in terms of an unique anionic current $\left<J_\phi^i\right>$ such that $\left<J_e^i\right>=e\alpha\,\left<J_\phi^i\right>$ and $\left<J_g^i\right>=g\hat{\beta}\,\left<J_\phi^i\right>$.

As for the statistical charge densities $\left<J_e^0\right>$ and $\left<J_g^0\right>$ we obtain clearly distinct results from the ones
obtained from the Chern-Simons model discussed in section~\ref{sec.int.CS}. From the magnetic Gauss' law given by the equation of motion~(\ref{EOM_C_0})
and the statistical electric charge density~(\ref{EOM_A_0}), considering the electromagnetic field definitions~(\ref{field_defs}),
we obtain the following relations between the external magnetic field $B$, the number of anyons
in the system $N$ and the statistical charge densities $\left<J_e^0\right>$ and $\left<J_g^0\right>$
\be
\left\{\ba{rcl}\left<J_g^0\right>&=&\displaystyle 2B+g\hat{\beta}\,N\\[3mm]
\left<J_e^0\right>&=&\displaystyle -\frac{2e\alpha}{g\hat{\beta}}\,B+e\alpha\,N\ea\right.\
 \Leftrightarrow\ 
\left\{\ba{rcl}B&=&\displaystyle -\frac{g\hat{\beta}}{4e\alpha}\left<J_e^0\right>+\frac{1}{4}\,\left<J_g^0\right>\\[3mm]
N&=&\displaystyle \frac{1}{2e\alpha}\,\left<J_e^0\right>+\frac{1}{2g\hat{\beta}}\,\left<J_g^0\right>\ea\right.
\lb{Je_0_Jg_0}
\ee 
Any combination of values for $\left<J_e^0\right>$ and $\left<J_g^0\right>$ obeying the above equalities can describe
a system containing $N$ anyons with an applied external orthogonal magnetic field of value $B$.
We will further discuss several particular fine-tuned configurations by the end of this section.
For now let us consider the standard assumption of the non-existence of magnetic charge, hence magnetic charge neutrality
of the planar system
\be
\left<J_g^0\right>\,=\,0\ \Rightarrow\ \left\{\ba{rcl}B&=&\displaystyle-\frac{g\hat{\beta}}{2e\alpha}\,\left<J_e^0\right>\\[5mm]N&=&\displaystyle\frac{1}{2e\alpha}\,\left<J_e^0\right>\ea\right.\ . 
\lb{J_e_0_0}
\ee
From these relations, for a given anyon number $N$ and coupling constants $e\alpha$ and $g\hat{\beta}$,
we obtain the following locking values for the external orthogonal magnetic field $B$, as well as the
respective value of the induced orthogonal electric field $\tilde{E}$~(\ref{sol_E_B}) and statistical electric charge density~(\ref{J_e_0_0})
\be
B\,=\,-\frac{g\hat{\beta}}{2}\,N\ ,\ \ \ \ \ \ \tilde{E}\,=\,\frac{e\alpha}{2}\,N\ ,\ \ \  \ \ \ \left<J_e^0\right>=2e\alpha\,N\ .
\lb{B_lock}
\ee 
These equations are the counterpart of the locking and vortex solutions given in equation~(\ref{ai})
for the Chern-Simons model discussed in section~\ref{sec.int.CS}. In the model we are presently discussing,
the internal gauge fields correspond to the pseudo-photon field $C_i$ and from the above solution for
$\tilde{E}$~(\ref{B_lock}) and the definition of the orthogonal electric field~(\ref{field_defs}) we obtain
the following vortex solution for the field components $C^i$
\be
C^i(\vb{r})=\frac{e\alpha}{2\pi}\,\epsilon^{ij}\int {dr'}^2\frac{r_j-r'_j}{|r-r'|^2}\phi^*\phi\ .
\lb{C_sol}
\ee
Similarly, from the lock equation for the magnetic field $B$~(\ref{B_lock}), we can write the formal equality
\be
\frac{g\hat{\beta}}{2\pi}\,\epsilon^{ij}\int {dr'}^2\frac{r_j-r'_j}{|r-r'|^2}\phi^*\phi=-A^i(\vb{r})\ .
\lb{A_sol}
\ee
However we stress that this is not a solution for the external gauge field, instead must be interpreted as that
the external gauge field induces vortex solutions in the planar system. We note that, if this model turns out to
actually have any relevance to Hall systems, these results are physically intuitive when interpreted in terms of the electromagnetic fields.
The interpretation is that the external magnetic field induces magnetic vortexes in the system with an orthogonal magnetic flux of opposite
direction (with value $g^*=-g\hat{\beta}$ per anyon) as expressed in~(\ref{A_sol}) which, in turn, create electric vortexes in the system
(the fractional charge of the anyons with value $e^*=e\alpha$ per anyon) as expressed in~(\ref{C_sol}) inducing an electric field orthogonal
to the system~(\ref{B_lock}). We recall that an (non-screened) electric charge usually creates an electric field, here the induced
field is due to the pseudo-photon gauge field ($\tilde{E}=\epsilon^{ij}\partial_iC_j/2$). As already discussed, such a mechanism is not
possible in planar system for Maxwell and Maxwell Chern-Simons theories, an orthogonal electric field is not present for these theories simply
because it is not described by these theories.

Of possible relevance to the Fractional Hall effect, from the results obtained so far and depending on the external fields configurations, we obtain the
following three distinct regimes:
\begin{itemize}
\item[\ \ I.] \textbf{no external fields applied},\\
$E^i=\tilde{E}^i=B=\tilde{B}=0$, we obtain the trivial solutions without any anyons in the system, no Hall current is present and no induced electromagnetic
fields are present, hence from the gauge field solutions $C_\mu=A_\mu=0$ we obtain 
\be
N=0\ \ \ ,\ \ \ \left<J^i_e\right>=0\ \ \ ,\ \ \ \tilde{E}=0\ \ \ ,\ \ \ \tilde{B}^i=0\ ;
\lb{pseudo_no_fields}
\ee
\item[\ II.] \textbf{only an orthogonal magnetic field is applied},\\
$E^i=0$ and $B=\epsilon^{ij}\partial_iA_j/2\neq 0$, the external magnetic field creates anyons in the planar system, no Hall current is present and it
is induced an orthogonal electric field, hence from the gauge field solutions $A_0=C_0=\partial_0C_i=0\ ,\ C_i=-e\alpha/(g\hat{\beta})\,A_i$ we obtain
\be
N=\frac{2}{g\hat{\beta}}\,|B|\ \ \ ,\ \ \ \left<J^i_e\right>=0\ \ \ ,\ \ \ \tilde{E}=-\frac{e\alpha}{g\hat{\beta}}\,B\ \ \ ,\ \ \ \tilde{B}^i=0\ ;
\lb{pseudo_B_field}
\ee
\item[III.] \textbf{both an external orthogonal magnetic field and a longitudinal electric field are applied},\\ $E^i=\partial_0 A_i-\partial_i A_0\neq 0$ and $B=\epsilon^{ij}\partial_iA_j/2\neq 0$, the external electric field induces an Hall current and a longitudinal magnetic field,
hence from the gauge field solutions $C_\mu=-e\alpha/(g\hat{\beta})\,A_\mu$ we obtain
\be
N=\frac{2}{g\hat{\beta}}\,|B|\ \ \ ,\ \ \ \left<J^i_e\right>=\hat{\sigma}_H^{ij}\,E_j\ \ \ ,\ \ \ \tilde{E}=-\frac{e\alpha}{g\hat{\beta}}\,B\ \ \ ,\ \ \ \tilde{B}^i=-\frac{e\alpha}{g\hat{\beta}}\,E^i\ ,
\lb{pseudo_E_B_fields}
\ee
with the Hall conductance given by $\hat{\sigma}_H^{ij}=e\alpha/(g\hat{\beta})\,\epsilon^{ij}$.
\end{itemize}

Therefore we obtained similar results to the Chern-Simons model discussed in section~\ref{sec.int.CS} with the equations~(\ref{pseudo_no_fields}),~(\ref{pseudo_B_field}) and~(\ref{pseudo_E_B_fields}) being the counterpart of
equations~(\ref{CS_no_fields}),~(\ref{CS_B_field}) and~(\ref{CS_E_B_fields}), respectively. However now both the vortex solutions
for the internal pseudo gauge field $C^i$~(\ref{C_sol}) correctly relates pseudo vectorial quantities as opposed to the vortex solution
for the internal gauge field $a^i$~(\ref{ai}) and the Hall conductance $\hat{\sigma}_H^{ij}$
transforms as a tensor under the discrete symmetries $P$ and $T$ due to being the product of
a pseudo-scalar ($e\alpha/(2g\hat{\beta})$) by a pseudo-tensor ($\epsilon^{ij}$) such that the Hall current is, correctly, defined as
a vectorial quantity. In addition are present both a non-null orthogonal electric field $\tilde{E}$ and longitudinal magnetic fields $\tilde{B}^i$
as expected from the electric and magnetic fluxes carried by the anyons in the system.

It is missing to properly justify the interpretation of the dimensionless coupling constants $\alpha$ and $\hat{\beta}$ as the average (in the
statistical sense) number of unit electric and magnetic flux vortex, respectively. In the next section we address this issue by
perturbing the solutions to the equations of motion.

\subsection{The Parameters $\alpha$ and $\hat{\beta}$ as the Average Number of unit Fluxes per Anyon}

In the standard Landau-Ginzburg Chern-Simons models presented in the introduction there are finite
energy solutions for the standard internal photon field $a$ which correspond to magnetic vortexes~\cite{CS_2,NO}
as expressed in equation~(\ref{vortex_CSa}). In the new model just derived with an internal pseudo-photon field $C$ we have, instead,
finite energy solutions corresponding to electric vortexes
\be
\phi=e^{\pm i\varphi}\ \ \ ,\ \ \ c^i=\pm\,\frac{e}{2\pi}\,\epsilon^{ij}\frac{r_j}{|\vb{r}|}\ .
\lb{vortex_C}
\ee
These vortexes are here considered to have one unit of electric charge $e$ and are written in
Cartesian coordinates. The existence of these configurations can be inferred directly from the solutions
for the equations of motion for the field components $C^i$ given in equation~(\ref{C_sol}).

Let us consider adding or removing one unit electric vortex to a planar system containing $N$ anyons by considering
a perturbation of the field components $C^i$ as expressed in~(\ref{C_sol}) by an amount $c^i$ as expressed in~(\ref{vortex_C}).
Hence the perturbed equation for the statistical electric charge density~(\ref{EOM_A_0}) is
\be
\epsilon^{ij}\partial_iC_j+\epsilon^{ij}\partial_ic_j=e\alpha N\ ,
\lb{e_charge}
\ee
with $N=\phi^*\phi$ and considering the value for the statistical electric charge density $\left<J^0_e\right>=2e\alpha\,N$~(\ref{B_lock}).
Integrating this equation we obtain the following solution for the $C^i$ field components
\be
C^i(\vb{r})=\frac{e}{2\pi}\,\left(\alpha\pm\frac{1}{N}\right)\,\epsilon^{ij}\int {dr'}^2\frac{r_j-r'_j}{|r-r'|^2}\phi^*\phi\ .
\ee
We can readily conclude that adding or removing an unit electric vortex to the system
shifts $\alpha$ by the amount $\pm 1/N$ such that the solution~(\ref{C_sol}) is retrieved by
redefining $\alpha\to \alpha'=\alpha\pm 1/N$ confirming the assumption for $\alpha$ being the average
number of unitary electric vortexes in the system, hence defined as
\be
\alpha=\frac{p}{N}\ \ \ ,\ \ \ p\,\in\,{\mathbb{N}}\ .
\lb{alpha_def}
\ee
Here $p$ is an integer representing the total number of electric vortexes in the system such that
$e\alpha\,N=e\,p$ is the total macroscopical statistical electric flux generated by the anyons in the planar system.

As already discussed we have assumed that the internal gauge field $a$ is non-dynamical, hence excluded
from the macroscopical theory. Let us show that, as already put forward in the previous section, although not being a dynamical field
the effect of the $a$ field is present in the model through the vortex solutions of the type
\be
\phi=e^{\pm i\varphi}\ \ \ ,\ \ \ a_i=\pm \frac{g}{2\pi}\,\epsilon^{ij}\,\frac{r_j}{|\vb{r}|}\ .
\lb{vortex_a}
\ee
These configurations have unit magnetic charge $g$ and are directly inferred from the lock condition on the external gauge field
components $A^i$ as expressed in~(\ref{A_sol}).

Hence adding or removing one of such vortexes to the system by perturbing the external field $A^i$ by an amount $a^i$ corresponding to the
unit vortex solution we obtain, from the magnetic Gauss' law equation~(\ref{EOM_C_0}),
\be
\epsilon^{ij}\partial_iA_j+\,\epsilon^{ij}\partial_ia_j=-g\hat{\beta}N\ ,
\ee
where, again, we have replaced $\phi^*\phi=N$ and consider null statistical magnetic charge density $\left<J^0_g\right>=0$~(\ref{J_e_0_0}).
Integrating this equation we obtain the locking condition for the external magnetic field
\be
\frac{g}{2\pi}\left(\hat{\beta}\pm\frac{1}{N}\right)\,\epsilon^{ij}\int {dr'}^2\frac{r_j-r'_j}{|r-r'|^2}\phi^*\phi=-A^i(\vb{r})\ .
\ee
Hence either condition~(\ref{A_sol}), or equivalently~(\ref{B_lock}), are retrieved by redefining $\hat{\beta}\to\hat{\beta}'\pm 1/N$ confirming
the assumption for $\hat{\beta}$ being the number of unit magnetic vortexes in the planar system defined as
\be
\hat{\beta}=\hat{1}\,\frac{q}{N}\ \ \ ,\ \ \ q\,\in\,{\mathbb{N}}\ .
\lb{beta_def}
\ee
Here $q$ is an integer representing the total number of magnetic vortexes in the planar system such that $g\hat{\beta}\,N=g\,q\,\hat{1}$
is the total magnetic flux due to the $N$ anyons. $\hat{1}$ is a pseudo-scalar number of unit modulus $|\hat{1}|=1$ and
has been introduced such that $\hat{\beta}$ is a pseudo-scalar quantity.

In addition we confirm that the magnetic vortexes are due to the internal gauge field $a$, the standard photon. The condensation
or screening effect is however not explained at all by this construction, as has been put forward in~\cite{planar}
a more fundamental description must be developed in order to understand and describe it (for instance, see~\cite{mpl} for a tentative
description employing canonical functional quantization). Here we have simply shown that
the existence of magnetic vortexes justifies our original assumptions when setting up the model.

In the next section we discuss the relation between the unit electric charge $e$ and unit magnetic charge $g$
assuming Dirac's quantization condition~\cite{Dirac,YW} and showing that, for the model derived here, it is equivalent to the
quantization of magnetic flux~\cite{London}.

\subsection{The Fractional Hall Conductance, Dirac's Quantization Condition and Quantization of Magnetic Flux}

So far we have not explicitly specified the relation between the unit electric charge $e$ and the unit magnetic charge $g$.
Pseudo-photon theory has been developed based in Dirac's theory of monopoles, hence we may assume Dirac quantization condition~\cite{Dirac}
\be
eg=n\,h\ ,
\ee
hence for $n=1$ we obtain, expressed in terms of the magnetic flux quantum $\Phi_0=h/2e$~\cite{London},
\be
\frac{e}{g}=\frac{e^2}{h}=\frac{e}{2\Phi_0}\ .
\lb{Dirac}
\ee
Therefore, from~(\ref{Je_i_Jg_i}), the Hall conductance is a pseudo-scalar quantity given by
\be
\hat{\sigma}_H=\frac{e\alpha}{g\hat{\beta}}=\frac{e^2}{h}\,\frac{\alpha}{\hat{\beta}}=\frac{e}{2\Phi_0}\,\frac{\alpha}{\hat{\beta}}\ ,
\lb{HA_AB}
\ee
being proportional to the ratio between the average electric flux and magnetic flux of the $N$ anyons in the planar system.

To further proceed let us note that experimentally the fractional charge of anyons $e^*$ is given by an odd fraction such
that $e^*=e/(2\bar{n}-1)$. Hence let us consider an adaptation from Haldane~\cite{Haldane} and Alperin~\cite{Alperin}
arguments (see also~\cite{Laughlin_3}) assuming that for an planar system, for each $2\bar{n}-1$ electrons there is an
electric vortex and for each $\bar{q}$ electrons there is a magnetic vortex. Given these assumptions we obtain
\be
\alpha=\frac{\frac{N}{2\bar{n}-1}}{N}=\frac{1}{2\bar{n}-1}\ \ \ \ ,\ \ \ \hat{\beta}=\frac{\frac{N}{\hat{\bar{q}}}}{N}=\frac{1}{\hat{\bar{q}}}\ ,
\lb{alpha_beta_def}
\ee
where $\hat{\bar{q}}$ is a pseudo-scalar number such that $|\hat{\bar{q}}|=|\bar{q}|$.
Within the model derived in this work we obtain straight forwardly the fractional electric flux $e^*=e\alpha=e/(2\bar{n}-1)$ and the fractional magnetic flux
$g\hat{\beta}=g/\hat{\bar{q}}$ carried by each anyon, as well as the fractional Hall conductance
\be
\hat{\sigma}_H=\frac{e}{2\Phi_0}\,\frac{\hat{\bar{q}}}{2\bar{n}-1}\ \ ,\ \ \ \hat{\bar{q}}\neq 0\in{\hat{\mathbb{N}}}\ ,\ \ \ \bar{n}\in{\mathbb{N}}\ ,
\lb{HA_H}
\ee
with the respective Landau level filling fraction $\nu_{\hat{\bar{q}},\bar{n}}=\hat{1}\alpha/\hat{\beta}=\bar{q}/(2\bar{n}-1)$.
Only in these last equations we explicitly considered $h$, in the remaining of this work we have considered natural units $h=1$.

We note that no justifications for
the odd denominator of the fractional charge neither for $\bar{q}$ and $2\bar{n}-1$ having no common prime factors are given
at the level of the model presented here. These characteristics are usually justified based in spin-statistics arguments.
Nevertheless this construction reproduces, at model level, the known results for Hall systems and the experimental verification of fractional charge
quantization of $1/(2\bar{n}-1)$ independently of $\bar{q}$~\cite{exp_charge}. Nevertheless it is relevant to note that within the model
presented so far the fluxes $e^*$ and $g^*$ are due to planar vortex, hence may correspond either to 4-dimensional charge or field configurations,
we will properly discuss this issue in section~\ref{sec.4D}. Also we have shown that for the model developed in
this work Dirac's quantization condition with $n=1$ is equivalent to the quantization of magnetic flux by the amount $\Phi_0=2e/h$.
However, as a final remark, we note that the extra orthogonal electric field and longitudinal magnetic fields are not present
in the standard Hall effect, hence this model does not describes this effect. If some suppression mechanism is considered, then it may
represent a standard Hall system. This may be the case, for example, of bi-layer Hall systems where these extra fields may become non-observable
contributing only to the inter-layer correlation function. We will not further develop here.

\subsection{Other Model Configurations}

So far, with respect to the value of the statistical charge densities $\left<J_e^0\right>$ and $\left<J_g^0\right>$, we have assumed global statistical magnetic charge
neutrality $\left<J_g^0\right>=0$. This case is a particular case for which the locking condition for the external magnetic field $B$ and the number of
anyons $N$ (composite electrons) in the system, as given in equation~(\ref{Je_i_Jg_i}), reproduces the Hall system constraints~(\ref{B_lock})
and vortex solutions for the internal gauge field~(\ref{C_sol}) maintaining the macroscopical magnetic charge null.

The constraints imposed on the statistical currents constitute a fine-tuning of the model and generally other fine-tuned values
for $\left<J_e^0\right>$ and $\left<J_g^0\right>$ can be considered. However, depending on the specific fine-tuning may be
harder, if not impossible, to interpret the solutions in terms of vortex solutions, hence the interpretation of the parameters
$\alpha$ and $\hat{\beta}$ may becomes unclear due to the electric and magnetic vortex solutions being mixed. We stress that
the choice $\left<J_g\right>=0$ was motivated by assuming null global magnetic charge. This discussion is not conclusive
until we properly discuss boundary conditions for the 4-dimensional embedding of the planar system in section~\ref{sec.4D}.
Let us postpone this discussion for a while and briefly list other particular fine-tuning choices.

\subsubsection{$\left<J_e^0\right>=0$, $\left<J_g^0\right>\neq 0$\label{sec.magnetic}}

As a particular case of interest we can consider statistical electric flux neutrality $\left<J_e^0\right>=0$ with non-neutral statistical magnetic charge $\left<J_g^0\right>\neq 0$. For this case similar constraints and solutions are obtained up to a negative sign, specifically we have
\be
\left<J_g^0\right>=2g\hat{\beta}N\ ,\ \ \left<J_e^0\right>=0\ \Rightarrow\ \left\{\ba{rcl}B&=&\displaystyle\frac{g\hat{\beta}}{2}\,N\\[5mm]\tilde{E}&=&\displaystyle-\frac{e\alpha}{2}\,N\ea\right.\ ,
\lb{magnetic}
\ee    
with the same Hall current $J_H(e)^i=\left<J_e^i\right>=\hat{\sigma}_H^{ij}E_j$ with the Hall conductance given in
equation~(\ref{HA_H}) and the vortex parameters $\alpha$ and $\hat{\beta}$ given in equation~(\ref{alpha_beta_def}).
Although the quantitative results closely match the case of statistical neutral magnetic charge $\left<J_g^0\right>=0$ with non-neutral
statistical electric charge $\left<J_e^0\right>=0$ discussed in detail in the previous sections, the interpretation is quite distinct:
the planar system, instead of a macroscopical statistical electric flux has a macroscopical statistical
magnetic flux.

\subsubsection{$\displaystyle \left<J_g^0\right>=+\frac{g\hat{\beta}}{e\alpha}\left<J_e^0\right>$\label{sec.selfdual}}

Other particular case is for self-dual charge configurations such that the equality $\left<J_g^0\right>=g\hat{\beta}/(e\alpha)\,\left<J_e^0\right>$
of the statistical charge densities is considered. From~(\ref{Je_0_Jg_0}) we obtain
\be
\left<J_g^0\right>=+\frac{g\hat{\beta}}{e\alpha}\left<J_e^0\right>\ \Rightarrow\ \left\{\ba{rcl}B&=&0\\[5mm]\left<J_e^0\right>&=&e\alpha\,N\,\ea\right.\ ,
\lb{selfdual}
\ee
also with $\tilde{E}=0$. Although anyons and an Hall current~(\ref{HA_H}) exist in the system as well as
the vortex solutions for the internal gauge field components $a_i$ and $c_i$ can be derived from the equations of motion for $A_0$ and $C_0$,
the external magnetic is null. Although by itself this configuration seems meaningless, when embedded in a 4-dimensional configurations
can be interpreted as describing a Meissner effect, although a magnetic field external to the system
may still be present, it does not penetrate the planar system. We will discuss this configuration later on from a 4-dimensional
thickened system perspective.

\subsubsection{$\displaystyle \left<J_g^0\right>=-\frac{g\hat{\beta}}{e\alpha}\left<J_e^0\right>$\label{sec.dual}}

For completeness let us also consider the particular case for anti-self-dual charge configurations such that the  equality
$\left<J_g^0\right>=-g\hat{\beta}/(e\alpha)\left<J_e^0\right>$ for the statistical charge densities is considered. From~(\ref{Je_0_Jg_0}) we obtain
\be
\left<J_g^0\right>=-\frac{g\hat{\beta}}{e\alpha}\left<J_e^0\right>\ \Rightarrow\ \left\{\ba{rcl}\left<J_e^0\right>&=&\displaystyle-\frac{2e\alpha}{g\hat{\beta}}B\\[5mm]N&=&0\ea\right.\ .
\lb{antiselfdual}
\ee    
We note that for this case ($N=\phi^*\phi=0$) the Lagrangian terms containing the anyonic fields are identically null.
Hence the parameters $\alpha$ and $\hat{\beta}$ are arbitrary and cannot be interpreted as the average vortex fluxes per (composite) particle
in the system: we are back to 3-dimensional pseudo-photon theory with statistical electric and magnetic currents
parameterized by the ratio $g\hat{\beta}/(e\alpha)$. Hence this configurations does not correspond to an Hall system either.

\section{4D Charges, Fields and Discrete Symmetries\lb{sec.4D}}

So far we have discussed only the statistical 3-dimensional charges. As already pointed out, a 3-dimensional charge
can correspond in 4-dimensions either to an orthogonal field or a charge configuration. In particular we note that 3-dimensional
vortexes usually only account for field flux orthogonal to the planar system and, when lifted back to 4-dimensions, can correspond
to flux tubes constituting either charge configurations either field configurations depending on the embedding of the 3-dimensional
manifold into the 4-dimensional manifold.

To distinguish between these two components of the 3-dimensional charge densities we note that the flux due to an electromagnetic field crossing
the planar system along the orthogonal direction $x^\perp$ has the same relative sign in both sides of the plane, while a charge configuration
generates an electromagnetic field of opposite directions at each side of the planar system, hence with a flux of opposite sign at each side
of the plane (with respect to $x^\perp$). Specifically we obtain that
\be
\ba{lll}
\mathrm{\bf field\ configurations}:&E_\perp^{(-)}=E_\perp^{(+)}\ ,&B_\perp^{(-)}=B_\perp^{(+)}\ ;\\[5mm]
\mathrm{\bf charge\ configurations}:&E_\perp^{(-)}=-E_\perp^{(+)}\ ,&B_\perp^{(-)}=-B_\perp^{(+)}\ .
\ea
\lb{field_charge_conf}
\ee
Here $x^i$ (with $i=1,2$) are the coordinates along the planar system and $x^\perp$ the spatial direction orthogonal
to the planar system. The superscript '$+$' sign stand for the values of the fields above the planar system ($x^\perp>0$) and the superscript '$-$'
sign stands for the values of the fields below the planar system ($x^\perp<0$). These two cases are pictured in figure~\ref{fig.1}.
\fig{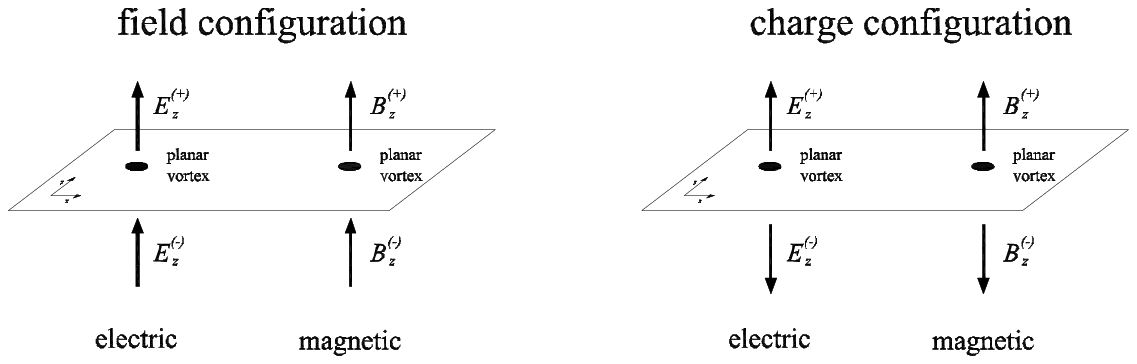}{$3+1$-dimensional field and charge configurations both for $2+1$-dimensional electric and magnetic vortex configurations for $x^1=x$, $x^2=y$ and $x^\perp=z$. Following equation~(\ref{field_charge_conf}), for field configurations the fields have symmetric boundary conditions with respect to the planar system ($E_z^{(-)}=E_z^{(+)}$ and $B_z^{(-)}=B_z^{(+)}$), while for charge configurations the fields have antisymmetric boundary conditions with respect to the planar system ($E_z^{(-)}=-E_z^{(+)}$ and $B_z^{(-)}=-B_z^{(+)}$).}{fig.1}
Hence we conclude that to properly identify whether the orthogonal fields induced by the planar system and the respective 3-dimensional statistical
charges correspond either to 4-dimentional charges, field configurations or a combination of both it is required to analyse their values
in two infinitesimal neighbourhoods along the orthogonal spatial direction above ($x^\perp>0$) and below ($x^\perp<0$) the planar system.

Next we analyse this construction based only on geometrical arguments and the symmetries of the planar system. Later on we will consider a more
realistic model by considering a thick planar system with two distinct boundaries and varying fields along the bulk.  

\subsection{Setting Relative Boundary Conditions}

To proceed let us recall that the orientation of the 3-dimensional manifold containing the planar system, as perceived in the
4-dimensional manifold from above and below the planar system, are reversed to each other. This is understood by noting that the exterior normal
to the system must be chosen either to point outwards or inwards of the system at the boundaries, hence if above the system it is pointing outwards,
below the system it must point downwards. This is a standard procedure in 4-dimensional electromagnetism and electrodynamics: whenever in the presence of a charged object is is required to choose a manifold orientation such that outwards and inward field fluxes are well defined.

To clarify this statement and properly set the relative orientations above and below the planar system,
as well as its embedding in the 4-dimensional manifold, let us consider an infinitesimal thickening of the planar system originally at $x^\perp=z=0$
by an amount $\delta_\perp$ along the orthogonal direction, hence obtaining two boundaries of the thickened planar system above and below $z=0$,
specifically at $z^{(\pm)}=\pm\delta_\perp/2$. We will refer to both these boundaries as upper sheet, $x^\perp=z^{(+)}$, and lower sheet,
$x^\perp=x^{(-)}$ (for details on relative boundary conditions in 3-dimensional systems see~\cite{Horava,PCT} and for the embedding of planar systems
see~\cite{planar}). Then, choosing a specific orientation for one of the boundaries of the thickened system imposes the inverse orientation in 
the other boundary, for a constant external magnetic field the transformation mapping each boundary into each other is
\be
\ba{lrcl}
{P\hspace{-2.5mm}\slash}_z: &z&\to& -z\ ;\\[5mm]
&B&\to& -B\ ;\\[5mm]
&\tilde{E}&\to&-\tilde{E}\ ;
\ea
\lb{PzO}
\ee
with the remaining fields and parameters of the model being invariant under this mapping, in particular the external electric field $E_i$, $N$, $e\alpha$
and $g\hat{\beta}$. We recall that the anti-symmetric tensor $\epsilon^{\mu\nu\lambda}$ (the Levi-Civita symbol)
is invariant under this transformation and that the pseudo-photon field is set by the equations of motion such that the transformation
for $\tilde{E}$ is fixed by the external magnetic field~(\ref{sol_E_B}). This transformation allows us to relate both boundaries of the thickened
planar system and infer that the embedding in the 4-dimensional manifold can be defined in the bounded space ${\mathbb{R}}^2\times[-\delta_\perp/2,\delta_\perp/2]/{P\hspace{-2.5mm}\slash}_z$ such that the planar system at $x^\perp=z=0$ is defined
in an orbifold plane~\cite{Horava,PCT}. This construction is pictured in figure~\ref{fig.2}.
\fig{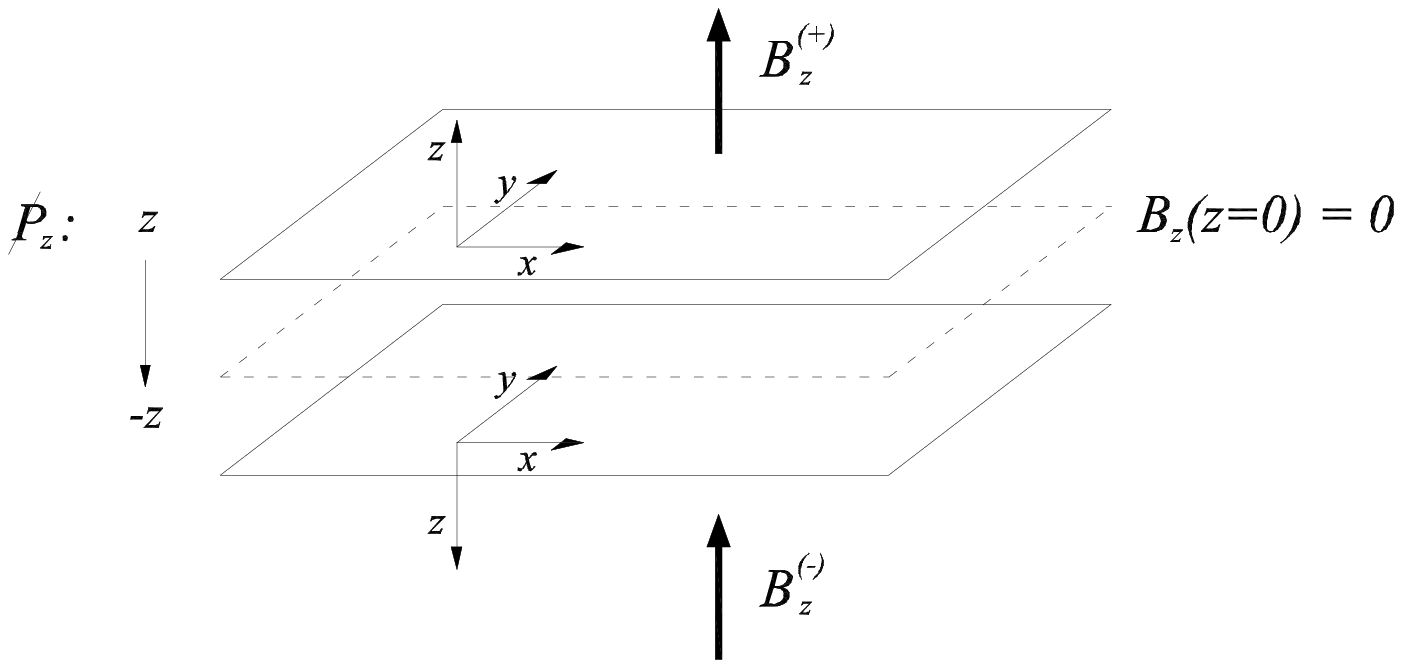}{$3+1$-dimensional embedding of a thick planar system with respective boundary sheet orientations given by the parity violating map ${P\hspace{-2.5mm}\slash}_z$~(\ref{PzO}).}{fig.2}
We note that configurations obeying the relative boundary conditions imposed by the map~(\ref{PzO}) explicitly break 4-dimensional parity $P$,
however this is a common feature of planar systems under external orthogonal fields. We will return to this discussion by the end of the manuscript.

\subsection{Estimating Statistical Currents in the 4D Manifold\lb{sec.estimative}}

From the above symmetry~(\ref{PzO}) we can estimate the statistical currents from the
perspective of the 4-dimensional world in which the 3-dimensional manifold (the planar system) is embedded.
In the following we will use the notation $\left<J_{e\mathrm{(4D)}}^\mu\right>$ and $\left<J_{g\mathrm{(4D)}}^\mu\right>$
(with brackets) for fully 4-dimensional current densities and $J_{e\mathrm{(4D)}}^\mu$ and $J_{g\mathrm{(4D)}}^\mu$ (without brackets) for the
respective 3-dimensional current densities such that these two distinct quantities are formally related by
\be
J_{e\mathrm{(4D)}}^\mu=\int^{+\infty}_{-\infty}\left<J_{e\mathrm{(4D)}}^\mu\right>\ \ ,\ \ J_{g\mathrm{(4D)}}^\mu=\int^{+\infty}_{-\infty}\left<J_{g\mathrm{(4D)}}^\mu\right>\ .
\lb{current_densities_rel}
\ee

As a first approach to estimate the current and charges from the perspective of the 4-dimensional world, we implement the map~(\ref{PzO}) by
considering the superposition of two copies of the planar systems, one slightly above and one slightly below $z=0$, related by this map, maintaining
however the same spatial orientation for both systems such that the map acts on the field configurations only leaving the spatial axis directions
unchanged. We consider the weight $1/2$ for each of the planar systems such that the current and charge densities are simply obtained by averaging
of the respective quantities at each system. 

We recall that the solutions for the equations of motion expressed in equation~(\ref{C}) are independent of any fine-tune in the statistical charge densities (or equivalently, independent of the fine-tune between $B$ and $N$). Hence the solution
for the vectorial components of the statistical currents given in equation~(\ref{Je_i_Jg_i}) is invariant under the map~(\ref{PzO}) and the
Hall current as perceived in the 4-dimensional manifold coincides with the 3-dimensional Hall current at each of the planar systems considered.
Specifically it is computed by averaging the Hall current in the upper and lower sheet of the system
\be
J_{H(e)\mathrm{(4D)}}^i=\frac{1}{2}\left(\left<J_{e\mathrm{(3D)}}^{i(+)}\right>+\left<J_{e\mathrm{(3D)}}^{i(-)}\right>\right)=\frac{e\alpha}{g\hat{\beta}}\,\epsilon^{ij}E_j\ .
\lb{Hall_current_4D_3D}
\ee
Consistently the same results are obtained for the magnetic vector current, $J_{H(g)\mathrm{(4D)}}^i=\epsilon^{ij}E_j/2$.

As for the electromagnetic charge densities and electromagnetic fields as perceived in the 4-dimensional manifold are obtained by half
the sum and half the differences of the respective 3-dimensional quantities in the upper and lower sheets  
\be
\ba{rcl}
J^0_{e\mathrm{(4D)}}&=&\displaystyle\frac{1}{2}\left(\left<J^{0(+)}_{e\mathrm{(3D)}}\right>+\left<J^{0(-)}_{e\mathrm{(3D)}}\right>\right)\ ,\\[5mm]
J^0_{g\mathrm{(4D)}}&=&\displaystyle\frac{1}{2}\left(\left<J^{0(+)}_{g\mathrm{(3D)}}\right>+\left<J^{0(-)}_{g\mathrm{(3D)}}\right>\right)\ ,\\[5mm]
E^\perp_{e\mathrm{(4D)}}&=&\displaystyle\frac{1}{2}\left(\frac{g\hat{\beta}}{2e\alpha}\left<J^{0(+)}_{e\mathrm{(3D)}}\right>-\frac{g\hat{\beta}}{2e\alpha}\left<J^{0(-)}_{e\mathrm{(3D)}}\right>\right)\ ,\\[5mm]
B^\perp_{e\mathrm{(4D)}}&=&\displaystyle\frac{1}{2}\left(\frac{1}{2}\left<J^{0(+)}_{g\mathrm{(3D)}}\right>-\frac{1}{2}\left<J^{0(-)}_{g\mathrm{(3D)}}\right>\right)\ .\\[5mm]
\ea
\label{4d_charges}
\ee
where we have taken in consideration the statistical charge densities expressions~(\ref{Je_0_Jg_0}), for a given external magnetic field $B$ and number of anyons (or electrons) in the system $N$. Further taking in consideration the orientation inversion transformation~(\ref{PzO}) we obtain the following upper and lower 3-dimensional charge densities
\be
\ba{rclcrcl}
\displaystyle\left<J_{g\mathrm{(3D)}}^{0(+)}\right>&=&\displaystyle+2B+g\hat{\beta} N&,&\left<J_{g\mathrm{(3D)}}^{0(-)}\right>&=&\displaystyle-2B+g\hat{\beta} N\ ,\\[5mm]
\displaystyle\left<J_{e\mathrm{(3D)}}^{0(+)}\right>&=&\displaystyle-\frac{2e\alpha}{g\hat{\beta}}B+e\alpha N&,&\left<J_{e\mathrm{(3D)}}^{0(-)}\right>&=&\displaystyle+\frac{2e\alpha}{g\hat{\beta}}B+e\alpha N\ ,
\ea
\ee
such that the charges and orthogonal fields evaluated from the perspective of the 4-dimensional manifold are
\be
\ba{rclcrcl}
J^0_{g\mathrm{(4D)}}&=&g\hat{\beta}\,N&,&B^\perp_{\mathrm{(4D)}}&=&B\ ,\\[5mm]
J^0_{e\mathrm{(4D)}}&=&e\alpha\,N&,&E^\perp_{\mathrm{(4D)}}&=&\tilde{E}\ .
\ea
\lb{4D_charges}
\ee
These results are independent of the specific fine-tuning for 3-dimensional statistical charge configurations, in particular whether we have
fine-tuned $\left<J_{g\mathrm{(4D)}}^0\right>=0$ with $\left<J_{e\mathrm{(4D)}}^0\right>\neq 0$ or $\left<J_e^0\right>=0$ with $\left<J_g\right>\neq 0$ at one of the boundaries. They depend only on the relative boundary conditions imposed by the map~(\ref{PzO}) applied to equation~(\ref{Je_0_Jg_0}), in particular
implying the following statistical charge fine-tuning between each boundary
\be
\left<J_{e{\mathcal(3D)}}^{0(-)}\right>=\frac{e\alpha}{g\hat{\beta}}\,\left<J_{g{\mathcal(3D)}}^{0(+)}\right>\ \ ,\ \ \ \left<J_{g{\mathcal(3D)}}^{0(-)}\right>=\frac{g\hat{\beta}}{e\alpha}\,\left<J_{e{\mathcal(3D)}}^{0(+)}\right> .
\lb{relative_3D_charges}
\ee
Given these results we conclude that the map~(\ref{PzO}) swaps the electric with the magnetic charge densities from the upper boundary
to the lower boundary.

It is relevant to note that due to these results not depending on the specific fine-tuning of the 3-dimensional statistical charge densities
the model presented here allows for a relatively large set of 3-dimensional configurations corresponding
to the same measurable effects in the 4-dimensional manifold. Although, at model level, this characteristic can be an advantage allowing to
incorporate distinct internal interpretations with respect to electric and magnetic charge location, from a fundamental point of view it clearly
fails to give an unique answer, in particular whether each composite particle (anyons) carries both electric and magnetic charge or independent electrical
and magnetically charged particles coexist in the system and whether these two kind of charged particles lay at distinct planes of the
thickened system. This discussion implies that to properly address this problem it should be considered a statistical description
of the thick system directly based in a microscopical theory/model accounting for the localization of charged excitations. We will not address this issue here.

We recall that the planar pseudo-photon action~(\ref{S_AC}) was deriving assuming all fields constant along the thickness
of the system $\delta_\perp$~\cite{planar}. In this section, mainly due to this characteristic, we have considered the superposition of two
planar systems related by the map~(\ref{PzO}). To implement this map considering one single thickened planar system it is required to consider
a 4-dimensional system with varying fields across the orthogonal direction. In particular it is required to obtain a bulk configuration in the range $x^\perp=z\in[-\delta_\perp/2,+\delta_\perp/2]$ that interpolates between the two boundaries
of the system such that the map~(\ref{PzO}) is obeyed for each two planes at $z$ and $-z$ in this interval. Specifically at the orbifold point of the map
$z=0$ we obtain
\be
x^\perp=z=0\ \ \Rightarrow\ \ \left\{\ba{l} B=-B=0\ ,\\[5mm]
\displaystyle\left<J_{e{\mathcal(4D)}}^{0(z=0)}\right>=\frac{e\alpha}{g\hat{\beta}}\,\left<J_{g{\mathcal(4D)}}^{0(z=0)}\right>\ .\ea\right.
\lb{orb_bc}
\ee
This planar configuration has already been discussed in subsection~\ref{sec.selfdual} being an allowed solution of the 3-dimensional
equations of motion with null external magnetic field $B$ and the desired relative fine-tuning between the electric and magnetic statistical
charges~(\ref{selfdual}). Here it is interpreted as that the thickened system has an effective null magnetic field at the orbifold plane due to
the relative boundary conditions, hence interpreted as a Meissner effect for the external magnetic field, it does not penetrate the planar system.

Next we formally define such a thickened planar system and explicitly compute the solutions of the equations of motion confirming the estimative
just derived from the map~(\ref{PzO}) based in geometrical arguments.

\subsection{Towards a more Realistic Model: Solutions of 4D Thick Equations of Motion\lb{sec.thick}}

In this section we consider the Lagrangian for a thick system of thickness $\delta_\perp$ defined in the
interval $z\in[-\delta_\perp/2,+\delta_\perp/2]$ and compute the respective 4-dimensional statistical currents from the equations of motion and the
current equations. We note that alternatively one could consider the system defined only between the orbifold plane and one of the boundaries
obtaining exactly the same results. 

We are considering the fields to have dependence on the $z$ coordinate although only with non-null components $\mu=0,x,y$ being
the orthogonal components of the fields assumed to be null
\be
A^\perp=C^\perp=\left<J^\perp_{e\mathrm{(4D)}}\right>=\left<J^\perp_{g\mathrm{(4D)}}\right>=0\ .
\ee
Also we note that here, the 4-dimensional statistical currents $\left<J^\mu_{e{\mathcal(4D)}}\right>(z)$ and $\left<J^\mu_{g{\mathcal(4D)}}\right>(z)$,
are evaluated for each value of the orthogonal coordinate $z$, hence should not be confused with the effective 3-dimensional
currents densities $J^\mu_{e{\mathcal(4D)}}$ and $J^\mu_{g{\mathcal(4D)}}$ expressed in equations~(\ref{Hall_current_4D_3D}) and~(\ref{4d_charges}) which are
obtained in the thin planar approximation (we will re-derive this quantities by the end of this section).

Then we consider the following Lagrangian inside the thickened system
\be
\ba{rcl}
z&\in&\displaystyle\left[-\frac{\delta_\perp}{2},+\frac{\delta_\perp}{2}\right]:\\[5mm]
{\mathcal{L}}^C_{\mathrm{(4D)}}&=&\displaystyle{\mathcal{L}}_{0\mathrm{(4D)}}^C-A_{\mu\mathrm{(4D)}}\,\left<J_{e\mathrm{(4D)}}^\mu\right>(z)-C_{\mu\mathrm{(4D)}} \left<J^\mu_{g\mathrm{(4D)}}\right>(z)\ ,\\[5mm]
{\mathcal{L}}_{0\mathrm{(4D)}}^C&=&{\mathcal{L}}_{\phi,C\mathrm{(4D)}}+{\mathcal{L}}_{\mathrm{(4D)}}\ ,\\[5mm]
{\mathcal{L}}_{\phi,C\mathrm{(4D)}}&=&\displaystyle-\phi_{\mathrm{(4D)}}^*\left[i\partial_0-e\,\alpha A_{0\mathrm{(4D)}}-g\hat{\beta} C_{0\mathrm{(4D)}}\right]\phi_\mathrm{(4D)}\\[5mm]
                      & &\displaystyle-\frac{1}{2\bar{m}}\phi_{\mathrm{(4D)}}^*\left[-i\vb{\nabla}-e\alpha\vb{A}_{\mathrm{(4D)}}-g\hat{\beta}\vb{C}_{\mathrm{(4D)}}\right]^2\phi_{\mathrm{(4D)}}\\[5mm]
                      & &+\mu_{\mathrm{(4D)}}\phi_{\mathrm{(4D)}}^*\phi_{\mathrm{(4D)}}-\lambda_{\mathrm{(4D)}}(\phi_{\mathrm{(4D)}}^*\phi_{\mathrm{(4D)}})^2\ ,
\ea
\lb{L_C_phi_2+1+1}
\ee
while considering the Lagrangian to be identically null outside the planar system
\be
\ba{rcl}
z&\in&\displaystyle{\mathbb{R}}/\left[-\frac{\delta_\perp}{2},+\frac{\delta_\perp}{2}\right]:\\[5mm]
{\mathcal{L}}^C_{\mathrm{(4D)}}&=&0\ .
\ea
\ee
Further assuming regularity of the remaining components of the gauge fields the $(2+1)+1$-splitting of the Lagrangian ${\mathcal{L}}_{\mathrm{(4D)}}$ is~\cite{planar}
\be
\ba{rcl}
{\mathcal{L}}_{\mathrm{(4D)}}&=&\displaystyle-\frac{1}{4}\,F_{\mu\nu\mathrm{(4D)}}F^{\mu\nu}_{\mathrm{(4D)}}+\frac{1}{4}\,G_{\mu\nu\mathrm{(4D)}}G^{\mu\nu}_{\mathrm{(4D)}}\\[5mm]
&&\displaystyle-\frac{1}{2}\,\partial^z A_{\mu\mathrm{(4D)}}\partial_z A^\mu_{\mathrm{(4D)}}+\frac{1}{2}\,\partial^z C_{\mu\mathrm{(4D)}}\partial_z C^\mu_{\mathrm{(4D)}}\\[5mm]
&&\displaystyle+\epsilon^{\mu\nu\lambda}\partial_z\left(A_{\mu\mathrm{(4D)}}\partial_\nu C_{\lambda\mathrm{(4D)}}\right)\ .
\ea
\lb{L_2+1+1}
\ee
We recall that, when the fields are regular, the derivative of the Chern-Simons term is a total derivative
not contributing to the bulk 4-dimensional equations of the motion. However this term does contribute to the boundary equations of motion,
to include this effect we integrate the Hopf term (the last term in the Lagrangian (\ref{L_2+1+1})) over the orthogonal
coordinate considering the same orientation for both boundaries obtaining the Chern-Simons terms at the boundaries
\be
\ba{l}
\displaystyle\int_{-\infty}^{+\infty}dz\,\epsilon^{\mu\nu\lambda}\partial_z\left(A_{\mu\mathrm{(4D)}}\partial_\nu C_{\lambda\mathrm{(4D)}}\right)=\\[5mm]
\ \ \ \ \ \ \ \ \displaystyle\int_{-\infty}^{+\infty}dz\,\left(\delta\left(z-\frac{\delta_\perp}{2}\right)+\delta\left(z+\frac{\delta_\perp}{2}\right)\right)\epsilon^{\mu\nu\lambda}\left(A_{\mu\mathrm{(4D)}}\partial_\nu C_{\lambda\mathrm{(4D)}}\right)\ .
\ea
\lb{L_3D}
\ee
Here we re-expressed this term using two Dirac delta functions such that these contributions
are interpreted as boundary current sources located at $z^{\pm}=\pm\delta_\perp/2$ for the thickened planar system.
We are taking the same orientation choice for both boundaries such that both boundaries contributions add
up, otherwise it would be required to consider a inversion of the $z$ axis direction across the thickness of the system. Here the
orientation inversion between both boundaries is modelled by the field solutions such that map~(\ref{PzO}) will still be obtained
by setting the appropriate boundary conditions without an explicit inversion of the $z$ axis. 

From the above Lagrangian~(\ref{L_C_phi_2+1+1}), considering the decomposition~(\ref{L_2+1+1}) and the boundary contribution~(\ref{L_3D})
we obtain the following equations of motion
\begin{align}
&\displaystyle\frac{\delta{\mathcal{L}}^C_{\mathrm(4D)}}{\delta C_{0\mathrm{(4D)}}}=0\ \Rightarrow\nonumber\\
&\displaystyle\hspace{0.5cm} \left<J^0_{g\mathrm{(4D)}}\right>(z)=\left(\delta\left(z-\frac{\delta_\perp}{2}\right)+\delta\left(z+\frac{\delta_\perp}{2}\right)\right)\epsilon^{ij}\partial_i A_{j\mathrm{(4D)}}+g\hat{\beta}\,\phi_{\mathrm{(4D)}}^*\phi_{\mathrm{(4D)}}\label{C_0_4D}-\partial_z\partial^z\,C^0_{\mathrm{(4D)}}\ ,\\[5mm]
&\displaystyle\frac{\delta{\mathcal{L}}^C_{\mathrm(4D)}}{\delta C_{i\mathrm{(4D)}}}=0\ \Rightarrow\nonumber\\
&\displaystyle\hspace{0.5cm}\left<J^i_{g\mathrm{(4D)}}\right>(z)=-\left(\delta\left(z-\frac{\delta_\perp}{2}\right)+\delta\left(z+\frac{\delta_\perp}{2}\right)\right)\epsilon^{ij}\left(\partial_0A_{j\mathrm{(4D)}}-\partial_jA_{0\mathrm{(4D)}}\right)\nonumber\\
&\displaystyle\hspace{3.5cm}-\frac{g\hat{\beta}}{\bar{m}}\left(e\alpha A^i_{\mathrm{(4D)}}+g\hat{\beta}C^i_{\mathrm{(4D)}}\right)\,\phi_{\mathrm{(4D)}}^*\phi_{\mathrm{(4D)}}-\partial_z\partial^zC^i_{\mathrm{(4D)}}\ ,\label{C_i_4D}\\[5mm]
&\displaystyle\frac{1}{\phi^*_{\mathrm{(4D)}}}\frac{\delta{\mathcal{L}}^C_{\mathrm(4D)}}{\delta \phi_{\mathrm{(4D)}}}=0\ \Rightarrow\nonumber\\
&\displaystyle\hspace{0.5cm}\left(e\alpha\,A_{0\mathrm{(4D)}}+g\hat{\beta}\,C_{0\mathrm{(4D)}}\right)-\frac{1}{2\bar{m}}\left(e\alpha \, A^i_{\mathrm{(4D)}}+g\hat{\beta}\,C^i_{\mathrm{(4D)}}\right)^2+\mu_{\mathrm{(4D)}}-2\lambda_{\mathrm{(4D)}}\,\phi_{\mathrm{(4D)}}^*\phi_{\mathrm{(4D)}}=0\label{phi_4D}\ .
\end{align}
and the statistical current equations
\begin{align}
&\displaystyle\frac{\delta{\mathcal{L}}^C_{\mathrm(4D)}}{\delta A_{0{\mathrm(4D)}}}=0\ \Rightarrow\nonumber\\
&\displaystyle\hspace{0.5cm}\displaystyle\left<J^0_{e\mathrm{(4D)}}\right>(z)=\left(\delta\left(z-\frac{\delta_\perp}{2}\right)+\delta\left(z+\frac{\delta_\perp}{2}\right)\right)\epsilon^{ij}\partial_i C_{j\mathrm{(4D)}}+e\alpha\,\phi_{\mathrm{(4D)}}^*\phi_{\mathrm{(4D)}}+\partial_z\partial^z\,A^0_{\mathrm{(4D)}}\ ,
\label{A_0_4D}\\[5mm]
&\displaystyle\frac{\delta{\mathcal{L}}^C_{\mathrm(4D)}}{\delta A_{i{\mathrm(4D)}}}=0\ \Rightarrow\nonumber\\
&\displaystyle\hspace{0.5cm}\left<J^i_{e\mathrm{(4D)}}\right>=-\left(\delta\left(z-\frac{\delta_\perp}{2}\right)+\delta\left(z+\frac{\delta_\perp}{2}\right)\right)\epsilon^{ij}\left(\partial_0 C_{j\mathrm{(4D)}}-\partial_jC_{0\mathrm{(4D)}}\right)+\partial_z\partial^z\,A^i_{\mathrm{(4D)}}\nonumber\\
&\displaystyle\hspace{3.5cm}-\frac{e\alpha}{\bar{m}}\left(e\alpha\,A^i_{\mathrm{(4D)}}+g\hat{\beta}\,C^i_{\mathrm{(4D)}}\right)\,\phi_{\mathrm{(4D)}}^*\phi_{\mathrm{(4D)}}\label{A_i_4D}\ .
\end{align}

To further proceed and solve these equations it is required to set the boundary conditions.
We will assume the upper and lower boundary solutions considered in the previous section as imposed by the map~(\ref{PzO})
such that in the upper half of the system, for $z>0$, the 3-dimensional magnetic statistical charge is null and in the lower
half of the system, for $z<0$, the 3-dimensional electric statistical charge is null according to~(\ref{relative_3D_charges})
\be
\left\{\ba{rcl}\left<J_{g\mathrm{(3D)}}^{0(+)}\right>&=&0\\[5mm]
\left<J_{e\mathrm{(3D)}}^{0(+)}\right>&=&e\alpha\,N\ea\right.\ \ \ ,\ \ \ 
\left\{\ba{rcl}\left<J_{e\mathrm{(3D)}}^{0(-)}\right>&=&g\hat{\beta}\,N\\[5mm]\left<J_{e\mathrm{(3D)}}^{0(-)}\right>&=&0\ea\right.\ \ .
\lb{boundary_charges}
\ee
We recall that this choice is not mandatory, up to an overall scaling constant, as discussed in the previous section the 4-dimesional charges
as well as Hall currents are independent of any fine-tuning of the 3-dimensional statistical charges as long as the parameters $\alpha$ and $\hat{\beta}$
are maintained fixed across the system thickness. However as we have already noticed this choice allows for a direct interpretation in terms of magnetic
and electric vortex solutions as expressed in equations~(\ref{C_sol}) and~(\ref{A_sol}). We further remark that, although when introducing the map~(\ref{PzO}) we have
considered relative opposite orientations at each boundary, here we are considered the same orientation at both boundaries to avoid an explicit inversion
of the $z$ axis, hence the field solutions will correspond to a modelling of the boundary conditions imposed by map~(\ref{PzO}) without an explicit
space orientation reversal, instead some of the fields will swap sign along the orthogonal direction reproducing the same effect of a orientation
reversal~\cite{footnote3}.

To implement the boundary conditions~(\ref{boundary_charges}) we are considering a coordinate field factorization $(2+1)+1$ (for the coordinates
$(z,x^\mu)$ with $\mu=0,1,2$) such that the 4-dimensional external fields, internal fields and currents in the thickened planar system factorize as
\be
\ba{rclcrcl}
A^0_{\mathrm{(4D)}}&=&k_A\,A^0_{\mathrm{(3D)}}(x^\mu)&,&A^i_{\mathrm{(4D)}}&=&f_A(z)\,A^i_{\mathrm{(3D)}}(x^\mu)\ ,\\[5mm]
C^0_{\mathrm{(4D)}}&=&k_C\,C^0_{\mathrm{(3D)}}(x^\mu)&,&C^i_{\mathrm{(4D)}}&=&f_C(z)\,C^i_{\mathrm{(3D)}}(x^\mu)\ ,\\[5mm]
\phi_{\mathrm{(4D)}}&=&k_\phi\phi_{\mathrm{(3D)}}\ ,
\ea
\lb{field_decomp}
\ee
where we introduced the normalization constants $k_A$, $k_C$ and $k_\phi$ which account for the thickness of the system
and the $z$ dependent functions $f_A(z)$ and $f_C(z)$ which account for the vectorial fields components variation across
the orthogonal coordinate $z$. We note that, for compatibility with the boundary condition~(\ref{boundary_charges}), $f_A(z)$
must be an antisymmetric function for which we impose the following boundary conditions
\be
f_A(z)=-f_A(-z)\ \ ,\ \ \ f_A\left(\pm\frac{\delta_\perp}{2}\right)=\pm \frac{1}{2}\ .
\lb{f_assymetry}
\ee
Hence we are assuming that the time-components $A^0$ and $C^0$ of the gauge fields, the vectorial components of the statistical currents $J_e^i$ and $J_g^i$
and the anyon field $\phi$ do not vary across the thickness of the system while the vectorial components of the gauge fields $A^i$ and $C^i$ do depend
on the orthogonal coordinate $x^\perp=z$. In particular we note that the planar spatial components $A^i$ of the external gauge field swap sign across
the system thickness. Again we remark that these field components are now interpreted as an effective bulk description of the external fields that
obey the map~(\ref{PzO}) such that the boundary orientation reversal are modelled by a regular field configuration, we stress that these do not
physically correspond to an inversion of the external gauge fields components, instead this construction, is interpreted as modelling the effective
external gauge field value inside the thickened system compatible with the boundary conditions imposed by~(\ref{PzO}).
We are further considering that the parameters $e\alpha$ and $g\hat{\beta}$ are constant along the orthogonal coordinate $z$.

Considering static and homogeneous solutions for the electromagnetic fields for each plane with fixed value of the coordinate $z$ we obtain the solution
\be
e\alpha A^i_{\mathrm{(4D)}}=-g\hat{\beta}C^i_{\mathrm{(4D)}}\ \ ,\ \ \mu_{\mathrm{(4D)}}=2\lambda_{\mathrm{(4D)}}\phi^*_{\mathrm{(4D)}}\phi_{\mathrm{(4D)}}\ .
\ee
Given the field decomposition~(\ref{field_decomp}) and demanding compatibility with the boundary conditions~(\ref{boundary_charges})
we obtain the following solutions
\be
\ba{l}
\displaystyle f_C(z)=f_A(z)\ \ ,\ \ k_\phi=\sqrt{\frac{1}{\delta_\perp}}\ \ ,\ \ \mu_{\mathrm{(3D)}}=\mu_{\mathrm{(4D)}}\ \ ,\ \ \lambda_{\mathrm{(3D)}}=\frac{\lambda_{\mathrm{(4D)}}}{\delta_\perp}\\[5mm]
\displaystyle C^i_{\mathrm{(3D)}}=-\frac{e\alpha}{g\hat{\beta}}A^i_{\mathrm{(3D)}}\ \ ,\ \ \phi^*_{\mathrm{(3D)}}\phi_{\mathrm{(3D)}}=N\ . 
\ea
\ee

Further considering the static external gauge field configuration
\be
A^0_{\mathrm(3D)}=x\,E_x+y\,E_y\ \ ,\ \ \ \vb{A}_{\mathrm(3D)}=\left(-\frac{yB}{2},\frac{xB}{2}\right)\ , 
\ee
a straight forward linear solution for these equations of motion is simply
\be
f_A(z)=f_C(z)=\frac{1}{\delta_\perp}\,z\ \ ,\ \ k_A=k_C=\frac{1}{2}
\ee
with the flux tube solutions for the pseudo-photon
\be
C^i_{\mathrm{(4D)}}(\vb{r},z)=\frac{e\alpha}{2\pi\,\delta_\perp}\,z\,\epsilon^{ij}\int {dr'}^2\frac{r_j-r'_j}{|r-r'|^2}\phi^*\phi\ ,
\lb{C_sol_thick}
\ee
and the lock equation for the external gauge field
\be
\frac{g\hat{\beta}}{2\pi\,\delta_\perp}\,z\,\epsilon^{ij}\int {dr'}^2\frac{r_j-r'_j}{|r-r'|^2}\phi^*\phi=-A^i_{\mathrm{(4D)}}(\vb{r},z)\ .
\lb{A_sol_thick}
\ee

Hence, from~(\ref{C_i_4D}) and~(\ref{A_i_4D}), are obtained the 4-dimensional Hall current density
\be
\left<J^i_{e\mathrm{(4D)}}\right>(z)=-\frac{e\alpha}{g\hat{\beta}}\left<J^i_{g\mathrm{(4D)}}\right>=\left\{\ba{lcl}
\displaystyle\frac{1}{2}\delta\left(z-\frac{\delta_\perp}{2}\right)\hat{\sigma}_H^{ij}\,E_j&,&\displaystyle z=\frac{\delta_\perp}{2}\\[5mm]
\displaystyle\frac{1}{2}\delta\left(z+\frac{\delta_\perp}{2}\right)\hat{\sigma}_H^{ij}\,E_j&,&\displaystyle z=-\frac{\delta_\perp}{2}\ea\right.
\ \ ,\ \ \hat{\sigma}_H^{ij}=\frac{e\alpha}{g\hat{\beta}}\,\epsilon^{ij}\ ,
\lb{J_i_e_4D}
\ee
being null for $z\neq \pm \delta_\perp/2$, hence localized at the boundaries $z=\pm \delta_\perp/2$.

As for the 4-dimensional statistical charge densities we obtain
\bea
\left<J^0_{e\mathrm{(4D)}}\right>(z)&=&\left\{\ba{lcl}
\displaystyle e\alpha\,N\,\left(\frac{1}{\delta_\perp}+\frac{1}{2}\delta\left(z-\frac{\delta_\perp}{2}\right)\right)&,&\displaystyle z=+\frac{\delta_\perp}{2}\\[5mm]
\displaystyle \frac{e\alpha}{\delta_\perp}\,N&,&\displaystyle z\in\left]-\frac{\delta_\perp}{2},\frac{\delta_\perp}{2}\right[\\[5mm]
\displaystyle +e\alpha\,N\,\left(\frac{1}{\delta_\perp}-\frac{1}{2}\delta\left(z+\frac{\delta_\perp}{2}\right)\right)&,&\displaystyle z=-\frac{\delta_\perp}{2}\ea\right.\lb{J_0_e_4D}\\[5mm]
\left<J^0_{g\mathrm{(4D)}}\right>(z)&=&\left\{\ba{lcl}
\displaystyle g\hat{\beta}\,N\,\left(\frac{1}{\delta_\perp}-\frac{1}{2}\delta\left(z-\frac{\delta_\perp}{2}\right)\right)&,&\displaystyle z=+\frac{\delta_\perp}{2}\\[5mm]
\displaystyle \frac{g\hat{\beta}}{\delta_\perp}\,N&,&\displaystyle z\in\left]-\frac{\delta_\perp}{2},\frac{\delta_\perp}{2}\right[\\[5mm]
\displaystyle g\hat{\beta}\,N\,\left(\frac{1}{\delta_\perp}+\frac{1}{2}\delta\left(z+\frac{\delta_\perp}{2}\right)\right)&,&\displaystyle z=-\frac{\delta_\perp}{2}\ea\right.\lb{J_0_g_4D}
\eea
being null outside of the planar system.

Hence we explicitly computed the 4-dimensional solutions for a thickened planar system for which the statistical 3-dimensional currents coincide with
the estimative given in section~\ref{sec.estimative} obeying the boundary conditions~(\ref{boundary_charges}), therefore justifying the
values obtained in equations~(\ref{Hall_current_4D_3D}) and~(\ref{4d_charges}). We note that the constants
$k_A$ and $k_C$ in the field solutions, as well as the normalization of the functions $f_A$ and $f_C$ have already been fine-tuned
to match the boundary conditions~(\ref{boundary_charges}).

Next we explicitly re-compute the values of the 3-dimensional current and charge densities and discussed the planar limit of the field solutions for
the thickened model just derived showing that, due to the relative boundary conditions imposed by the map~(\ref{PzO}), they do not coincides with the
model discussed in section~\ref{sec.electric}, instead both non-null statistical electric and magnetic charges must be present and the effective
external magnetic field is null in the plane $z=0$ due to the orbifold relation imposed by map~(\ref{PzO}).

\subsection{(Re-)Computing Statistical Currents and the Thin Planar Approximation}

The 3-dimensional vector current and charge densities, as perceived from the perspective of the 4-dimensional manifold, are computed simply by integrating
over the orthogonal coordinate~(\ref{current_densities_rel}), hence from~(\ref{J_i_e_4D}),~(\ref{J_0_e_4D}) and~(\ref{J_0_g_4D}) we straight forwardly obtain
\bea
J^i_{H(e)\mathrm{(4D)}}&=&-\frac{e\alpha}{g\hat{\beta}}\,J^i_{H(g)\mathrm{(4D)}}=\int_{-\infty}^{+\infty}\left<J^i_{e\mathrm{(4D)}}\right>=\hat{\sigma}^{ij}_H\,E_j\ \ ,\ \ \hat{\sigma}_H^{ij}=\frac{e\alpha}{g\hat{\beta}}\,\epsilon^{ij}\ ,\\
J^0_{e\mathrm{(4D)}}&=&\int_{-\infty}^{+\infty}\left<J^0_{e\mathrm{(4D)}}\right>=\frac{e\alpha}{\delta_\perp}\,N\,z\Bigl|^{+\frac{\delta_\perp}{2}}_{-\frac{\delta_\perp}{2}}=e\alpha\,N\ ,\\
J^0_{g\mathrm{(4D)}}&=&\int_{-\infty}^{+\infty}\left<J^0_{g\mathrm{(4D)}}\right>=\frac{g\hat{\beta}}{\delta_\perp}\,N\,z\Bigl|^{+\frac{\delta_\perp}{2}}_{-\frac{\delta_\perp}{2}}=g\hat{\beta}\,N\ ,
\eea
Here we are using the same notation of equation~(\ref{current_densities_rel}), the quantities $J^\mu_{e\mathrm{(4D)}}$ and $J^\mu_{g\mathrm{(4D)}}$
(without brakets) represent the 3-dimensional current and charge densities as perceived in the 4-dimensional manifold. As for the 3-dimensional
current and charge densities for the planar system discussed in section~\ref{sec.estimative} can be computed by integrating the respective
4-dimensional quantities over the upper half and lower half of the thick system
\bea
\left<\bar{J}^{i(+)}_{e\mathrm{(3D)}}\right>&=&\left<\bar{J}^{i(-)}_{e\mathrm{(3D)}}\right>=\int_{0}^{+\infty}\left<J^i_{e\mathrm{(4D)}}\right>=\frac{1}{2}\hat{\sigma}^{ij}_H\,E_j\ ,\\[5mm]
\left<\bar{J}^{i(+)}_{g\mathrm{(3D)}}\right>&=&\left<\bar{J}^{i(-)}_{g\mathrm{(3D)}}\right>=\int^{0}_{-\infty}\left<J^i_{g\mathrm{(4D)}}\right>=\frac{g\hat{\beta}}{2e\alpha}\hat{\sigma}^{ij}_H\,E_j\ ,\\[5mm]
\left<\bar{J}^{0(+)}_{e\mathrm{(3D)}}\right>&=&\int_{0}^{+\infty}\left<J^0_{e\mathrm{(4D)}}\right>=\frac{e\alpha}{\delta_\perp}\,N\,z\Bigl|^{\frac{\delta_\perp}{2}}_{0}+\frac{e\alpha}{2}\,N=e\alpha\,N\ ,\\[5mm]
\left<\bar{J}^{0(-)}_{e\mathrm{(3D)}}\right>&=&\int^{0}_{-\infty}\left<J^0_{e\mathrm{(4D)}}\right>=\frac{e\alpha}{\delta_\perp}\,N\,z\Bigl|_{-\frac{\delta_\perp}{2}}^{0}-\frac{e\alpha}{2}\,N=0\ ,\\[5mm]
\left<\bar{J}^{0(+)}_{g\mathrm{(3D)}}\right>&=&\int_{0}^{+\infty}\left<J^0_{g\mathrm{(4D)}}\right>=\frac{g\hat{\beta}}{\delta_\perp}\,N\,z\Bigl|^{\frac{\delta_\perp}{2}}_{0}-\frac{g\hat{\beta}}{2}\,N=0\ ,\\[5mm]
\left<\bar{J}^{0(-)}_{g\mathrm{(3D)}}\right>&=&\int^{0}_{-\infty}\left<J^0_{g\mathrm{(4D)}}\right>=\frac{g\hat{\beta}}{\delta_\perp}\,N\,z\Bigl|_{-\frac{\delta_\perp}{2}}^{0}+\frac{g\hat{\beta}}{2}\,N=g\hat{\beta}\,N\ ,
\eea
We stress that these quantities do not exactly match the ones computed in section~\ref{sec.estimative} where a superposition of two
planar systems related by map~(\ref{PzO}) were considered such that the current and charge densities were computed by averaging the
same quantities for both these systems. This construction was required due to the two planar systems considered in section~\ref{sec.estimative}
being obtained by dimensional reduction of two superimposed thick systems of thickness $\delta_\perp$ with constant fields along the orthogonal coordinate.
The thick model just derived represents a single system of thickness $\delta_\perp$ having varying fields across the orthogonal coordinate such that
the map~(\ref{PzO}) is valid for every two planes at $z$ and $-z$ along the orthogonal coordinate. Hence the bar quantities just computed and
the respective quantities in section~\ref{sec.estimative} are related by a factor of two
\be
\left<J^{\mu(\pm)}_{e\mathrm{(3D)}}\right>=2\left<\bar{J}^{\mu(\pm)}_{e\mathrm{(3D)}}\right>\ \ ,\ \ \left<J^{\mu(\pm)}_{g\mathrm{(3D)}}\right>=2\left<\bar{J}^{\mu(\pm)}_{g\mathrm{(3D)}}\right>\ ,
\ee
being the values given above particularized to the boundary conditions~(\ref{boundary_charges}).

Once we computed the statistical current and charge densities for the thick system we can return to a planar description
of the system. This can be achieved by taking the thin limit approximation $\delta_\perp\to 0$  such that $f_C(z)=f_A(z)\to H(z)$,
being $H(z)$ the Heaviside function ($H(z<0)=-1$, $H(z>0)=+1$). The most interesting result is that
the \textit{thick vortex} solutions, consistently with the 4-dimensional charges interpretation for the model~(\ref{4D_charges}),
correspond now to flux tubes with variable flux along the $z$ coordinate having a positive flux for $z>0$ and a negative flux for $z<0$ as expected from
a charge-like configuration. We also remark that at $z=0$ the flux is null as expected from the map~(\ref{relative_3D_charges}) for which the
planar system configuration corresponds to the self-dual charge configuration discussed in subsection~\ref{sec.selfdual}.
In this thin limit exists a discontinuity at the planar system (the plane at $z=0$) such that the vortex solution and magnetic field
locking condition are
\be
\ba{l}
\displaystyle C^i_{\mathrm{(2+1+1D)}}(\vb{r},z)=e\alpha\,H(z)\,\epsilon^{ij}\int {dr'}^2\frac{r_j-r'_j}{|r-r'|^2}\phi^*\phi\ ,\\[5mm] 
\displaystyle g\hat{\beta}\,H(z)\,\epsilon^{ij}\int {dr'}^2\frac{r_j-r'_j}{|r-r'|^2}\phi^*\phi=-A^i_{\mathrm{(2+1+1D)}}(\vb{r},z)\ .
\ea
\lb{flux_tubes}
\ee
and the 3-dimensional charges coincide with the self-dual configuration of the planar system with a null external magnetic field
\be
\left<J^0_{g\mathrm{(3D)}}\right>(z=0)=2g\hat{\beta}\,N\ \ ,\ \ \ \left<J^0_{e\mathrm{(3D)}}\right>(z=0)=2e\alpha\,N\ .
\ee
Once more we recall that the value of the external magnetic field is interpreted as the effective field at the plane containing the system as already discussed.

Hence we have just derived the solutions of the equations of motion for a thickened planar system retrieving, in the thin limit approximation,
a consistent description of charges and fields. We stress again that the results just derived are valid for any fine-tuning of the 3-dimensional
charge configurations across the orthogonal direction of the thickened system and that, although the 4-dimensional measurable effects are the same
for all these configurations, the physical interpretation with respect to the internal interactions are distinct.

As a particular example we note that the fine-tuning considered in this section is interpreted as that electric and magnetic charged excitations 
lay in distinct planes along the orthogonal direction to the thickened system, specifically the 3-dimensional electric and magnetic charges lay
at opposite boundaries.
As another particular example, taking the self-dual configuration for all values of $z$,
$\left<J^0_{g\mathrm{(3D)}}\right>(z)=e\alpha/(g\hat{\beta})\left<J^0_{e\mathrm{(3D)}}\right>(z)$ we conclude
that it corresponds to the coexistence of electric and magnetic charged configurations at the same plane of the thickened system and that
the external magnetic field is null at the plane of the system such that a Meissner effect is present. Clearly these two configurations
correspond to distinct interactions between the electric and magnetic excitations. This is not completely unexpected, usually a macroscopical
description of any system has reduced degrees of freedom. Nevertheless we can conclude from this discussion that a planar description does
not allow to fully describe the system, instead it may be required a thickened model to properly describe the internal interactions.

\subsection{3D and 4D Discrete Symmetries $P$, $T$ and $C$}

So far we have not explicitly analysed the parity transformation for the model and configurations discussed in this article.
As already mentioned 4-dimensional parity (in this section we use the notation $P_{\mathrm{(4D)}}$) is explicitly violated due to the
external magnetic field. As for 3-dimensional parity
(in this section we use the notation $P_{\mathrm{(3D)}}$) is conserved at the planar system. This result can be verified as usual
by explicitly showing that the equations of motion correctly transform under the parity transformation. We also discuss
time inversion ($T_{\mathrm{(4D)}}$ and $T_{\mathrm{(3D)}}$) and charge conjugation ($C_{\mathrm{(4D)}}$ and $C_{\mathrm{(3D)}}$).

4-dimensional parity inverts the spatial coordinates $(x,y,z)$, hence corresponding to an inversion of spatial orientation,
while time inversion reverses the time coordinate direction $t$. The several vector and pseudo-vector fields and currents and
scalar and pseudo-scalar charge densities are mapped according to the following transformations~\cite{action,Jackson}:
\be
\ba{lrclclrcl}
P_{\mathrm{(4D)}}:&x^i&\to&-x^i&\ &T_{\mathrm{(4D)}}:&t&\to&-t\\[5mm]
&A_0&\to&+A_0&&&A_0&\to&-A_0\\[5mm]
&A_i&\to&-A_i&&&A_i&\to&+A_i\\[5mm]
&C_0&\to&-C_0&&&C_0&\to&+C_0\\[5mm]
&C_i&\to&+C_i&&&C_i&\to&-C_i\\[5mm]
&E^i_{\mathrm{(4D)}}&\to&-E^i_{\mathrm{(4D)}}&&&E^i_{\mathrm{(4D)}}&\to&+E^i_{\mathrm{(4D)}}\\[5mm]
&B^i_{\mathrm{(4D)}}&\to&+B^i_{\mathrm{(4D)}}&&&B^i_{\mathrm{(4D)}}&\to&-B^i_{\mathrm{(4D)}}\\[5mm]
&\rho_{e\mathrm{(4D)}}&\to&+\rho_{e\mathrm{(4D)}}&&&\rho_{e\mathrm{(4D)}}&\to&+\rho_{e\mathrm{(4D)}}\\[5mm]
&J^i_{e\mathrm{(4D)}}&\to&-J^i_{e\mathrm{(4D)}}&&&J^i_{e\mathrm{(4D)}}&\to&-J^i_{e\mathrm{(4D)}}\\[5mm]
&\rho_{g\mathrm{(4D)}}&\to&-\rho_{g\mathrm{(4D)}}&&&\rho_{g\mathrm{(4D)}}&\to&-\rho_{g\mathrm{(4D)}}\\[5mm]
&J^i_{g\mathrm{(4D)}}&\to&+J^i_{g\mathrm{(4D)}}&&&J^i_{g\mathrm{(4D)}}&\to&+J^i_{g\mathrm{(4D)}}
\ea
\ee
with $i=x,y,z$. As for charge conjugation $C_{\mathrm{(4D)}}$ it swaps the sign of all the 4-fields and 4-currents.
Next we analyse the effect of these symmetries in the planar quantities by assuming that at each plane of fixed $z$
their value correspond to the value of the respective 4-dimensional quantities for that specific value of the coordinate $z$.

Violation of $P_{\mathrm{(4D)}}$ can be explicitly checked by the transformation of the Hall current definitions~(\ref{Je_i_Jg_i}) and charge definitions~(\ref{Je_0_Jg_0}), noting that under 4-dimensional parity the parameter $\hat{\beta}$ swaps sign while the parameter $\alpha$ is
unchanged we obtain
\be
\ba{llcl}
P_{\mathrm{(4D)}}:&
\displaystyle +\rho_{e}=-\frac{e\alpha}{g\hat{\beta}}\,B+e\alpha\,N&\to&\displaystyle+\rho_{e}=+\frac{e\alpha}{g\hat{\beta}}\,B+e\alpha\,N\\[5mm]
 &\displaystyle+\rho_{g}=+B+g\hat{\beta}\,N&\to&\displaystyle-\rho_{g}=+B-g\hat{\beta}\,N\\[5mm]
 &\displaystyle+J_{H(e)}^i=+\frac{e\alpha}{2g\hat{\beta}}\,\epsilon^{ij}E_j&\to&\displaystyle-J_{H(e)}^i=+\frac{e\alpha}{2g\hat{\beta}}\,\epsilon^{ij}E_j\\[5mm]
 &\displaystyle+J_{H(g)}^i=-\frac{1}{2}\epsilon^{ij}E_j&\to&\displaystyle+J_{H(g)}^i=+\frac{1}{2}\,\epsilon^{ij}E_j
\ea
\ee
hence these expression are not invariant explicitly violating parity.

Time inversion $T_{\mathrm{(4D)}}$ is conserved, this can be explicitly checked by noting that the parameter $\hat{\beta}$ swaps sign while
the parameter $\alpha$ remains unchanged and considering the explicit transformations for the Hall currents~(\ref{Je_0_Jg_0})
and the charge definitions~(\ref{Je_0_Jg_0})
\be
\ba{llcl}
T_{\mathrm{(4D)}}:&
 \displaystyle+\rho_{e}=-\frac{e\alpha}{g\hat{\beta}}\,B+e\alpha\,N&\to&\displaystyle+\rho_{e}=-\frac{e\alpha}{g\hat{\beta}}\,B+e\alpha\,N\\[5mm]
 &\displaystyle+\rho_{g}=+B+g\hat{\beta}\,N&\to&\displaystyle-\rho_{g}=-B-g\hat{\beta}\,N\\[5mm]

 &\displaystyle+J_{H(e)}^i=+\frac{e\alpha}{2g\hat{\beta}}\,\epsilon^{ij}E_j&\to&\displaystyle-J_{H(e)}^i=-\frac{e\alpha}{2g\hat{\beta}}\,\epsilon^{ij}E_j\\[5mm]
 &\displaystyle+J_{H(g)}^i=-\frac{1}{2}\epsilon^{ij}E_j&\to&\displaystyle+J_{H(g)}^i=-\frac{1}{2}\,\epsilon^{ij}E_j
\ea
\lb{T_4D}
\ee
hence time inversion is conserved. 

Charge conjugation $C_{\mathrm{(4D)}}$ is trivially conserved, all quantities swap sign and the definitions of charges and currents
consistently change the sign of both the right and left hand side. 

As for $P_{\mathrm{(4D)}}T_{\mathrm{(4D)}}$ it is also violated, we note that this discrete transformation swaps the sign of both electric and magnetic fields
not changing the sign for both electric and magnetic currents and charge densities, hence the transformation of the electric Hall current explicitly violates
this symmetry. This result may also be directly inferred from violation of parity and conservation of time inversion. The same arguments follow for $P_{\mathrm{(4D)}}C_{\mathrm{(4D)}}$. To show that $T_{\mathrm{(4D)}}C_{\mathrm{(4D)}}$ is conserved it is enough to note that each of the symmetries $T_{\mathrm{(4D)}}$ and $C_{\mathrm{(4D)}}$ is independently conserved.

For last also $PCT$ is violated due to $P$ violation and $PT$ conservation.
We recall that $PCT$ violation is in quantum field theories (in particular QED) an unusual occurrence.
However we note that for the model being discussed it is simply
due to parity violation which, in turn is due to the external orthogonal magnetic field applied to a planar configuration of the system,
in simple terms the physics is constraint to a plane (or thickened plane) such that the full 4-dimensional symmetry no longer apply.

Next we analyse the 3-dimensional discrete symmetries for fixed values of the coordinate $z$ ($z=0$ when a planar system is considered)
concluding that these symmetries are actually conserved.
In 3-dimensions a inversion of space orientation correspond to the inversion of one of the axis (the inversion of both $x$ and $y$ is equivalent to
a planar rotation). Hence we choose to invert the $y$ coordinate such that $P_{\mathrm{(3D)}}$ and $T_{\mathrm{(3D)}}$ are defined as~\cite{PCT}:
\be
\ba{lrclclrcl}
P_{\mathrm{(3D)}}:&y&\to&-y&\ &T_{\mathrm{(3D)}}:&t&\to&-t\\[5mm]
&A_0&\to&+A_0&&&A_0&\to&-A_0\\[5mm]
&A_x&\to&+A_x&&&A_i&\to&+A_i\\[5mm]
&A_y&\to&-A_y&&&\\[5mm]
&C_0&\to&-C_0&&&C_0&\to&+C_0\\[5mm]
&C_x&\to&-C_x&&&C_i&\to&-C_i\\[5mm]
&C_y&\to&+C_y&&&\\[5mm]
&E^x_{\mathrm{(3D)}}&\to&+E^x_{\mathrm{(3D)}}&&&E^i_{\mathrm{(3D)}}&\to&+E^i_{\mathrm{(3D)}}\\[5mm]
&E^y_{\mathrm{(3D)}}&\to&-E^y_{\mathrm{(3D)}}&&&\\[5mm]
&\tilde{E}_{\mathrm{(3D)}}&\to&+\tilde{E}_{\mathrm{(3D)}}&&&\tilde{E}_{\mathrm{(3D)}}&\to&+\tilde{E}_{\mathrm{(3D)}}\\[5mm]
&\tilde{B}^x_{\mathrm{(3D)}}&\to&-\tilde{B}^x_{\mathrm{(3D)}}&&&\tilde{B}^i_{\mathrm{(3D)}}&\to&-\tilde{B}^i_{\mathrm{(3D)}}\\[5mm]
&\tilde{B}^y_{\mathrm{(3D)}}&\to&+\tilde{B}^y_{\mathrm{(3D)}}&&&\\[5mm]
&B_{\mathrm{(3D)}}&\to&-B_{\mathrm{(3D)}}&&&B_{\mathrm{(3D)}}&\to&-B_{\mathrm{(3D)}}\\[5mm]
&\rho_{e\mathrm{(3D)}}&\to&+\rho_{e\mathrm{(3D)}}&&&\rho_{e\mathrm{(3D)}}&\to&+\rho_{e\mathrm{(3D)}}\\[5mm]
&J^x_{e\mathrm{(3D)}}&\to&+J^x_{e\mathrm{(3D)}}&&&J^i_{e\mathrm{(3D)}}&\to&-J^i_{e\mathrm{(3D)}}\\[5mm]
&J^y_{e\mathrm{(3D)}}&\to&-J^y_{e\mathrm{(3D)}}&&&\\[5mm]
&\rho_{g\mathrm{(3D)}}&\to&-\rho_{g\mathrm{(3D)}}&&&\rho_{g\mathrm{(3D)}}&\to&-\rho_{g\mathrm{(3D)}}\\[5mm]
&J^x_{g\mathrm{(4D)}}&\to&-J^x_{g\mathrm{(4D)}}&&&J^i_{g\mathrm{(3D)}}&\to&+J^i_{g\mathrm{(3D)}}\\[5mm]
&J^y_{g\mathrm{(4D)}}&\to&+J^y_{g\mathrm{(4D)}}&&&
\ea
\ee
and 3-dimensional charge conjugation $C_{\mathrm{(3D)}}$ swaps the signs of all fields and charges.

Noting that under 3-dimensional parity the parameter $\hat{\beta}$ swaps sign while the parameter $\alpha$
is unchanged we obtain the following transformation of the Hall currents~(\ref{Je_i_Jg_i}) and charges~(\ref{Je_0_Jg_0})
under parity:
\be
\ba{llcl}
P_{\mathrm{(3D)}}:&
\displaystyle +\rho_{e}=-\frac{e\alpha}{g\hat{\beta}}\,B+e\alpha\,N&\to&\displaystyle+\rho_{e}=-\frac{e\alpha}{g\hat{\beta}}\,B+e\alpha\,N\\[5mm]
 &\displaystyle+\rho_{g}=+B+g\hat{\beta}\,N&\to&\displaystyle-\rho_{g}=-B-g\hat{\beta}\,N\\[5mm]
 &\displaystyle+J_{H(e)}^x=+\frac{e\alpha}{2g\hat{\beta}}\,\epsilon^{xy}E_y&\to&\displaystyle+J_{H(e)}^x=+\frac{e\alpha}{2g\hat{\beta}}\,\epsilon^{xy}E_y\\[5mm]
 &\displaystyle+J_{H(e)}^y=+\frac{e\alpha}{2g\hat{\beta}}\,\epsilon^{yx}E_x&\to&\displaystyle-J_{H(e)}^y=-\frac{e\alpha}{2g\hat{\beta}}\,\epsilon^{yx}E_x\\[5mm]
 &\displaystyle+J_{H(g)}^x=-\frac{1}{2}\epsilon^{xy}E_y&\to&\displaystyle-J_{H(g)}^x=-\frac{1}{2}\,\epsilon^{xy}E_y\\[5mm]
 &\displaystyle+J_{H(g)}^y=-\frac{1}{2}\epsilon^{yx}E_x&\to&\displaystyle+J_{H(g)}^y=+\frac{1}{2}\,\epsilon^{yx}E_x
\ea
\ee
hence we conclude that 3-dimensional parity is conserved.

3-dimensional time inversion $T_{\mathrm{(3D)}}$ acts in the fields exactly as the respective 4-dimensional symmetry~(\ref{T_4D}) being
also conserved. Finally 3-dimensional charge conjugation $C_{\mathrm{(3D)}}$ is also conserved, it acts in the fields and charges by reversing
all signs such that all the definitions are trivially invariant under this symmetry. Since the three discrete symmetries are conserved in
3-dimensions all the other possible combinations $P_{\mathrm{(3D)}}T_{\mathrm{(3D)}}$, $P_{\mathrm{(3D)}}C_{\mathrm{(3D)}}$, $T_{\mathrm{(3D)}}C_{\mathrm{(3D)}}$ and $P_{\mathrm{(3D)}}C_{\mathrm{(3D)}}T_{\mathrm{(3D)}}$ are also conserved symmetries.

We summarize the results obtained with respect to the discrete symmetries:\\
\begin{center}
\begin{tabular}{lcc}\hline
 & 4-dimensions & 3-dimensions\\[2mm]\hline\hline
$P$&violated&conserved\\[5mm]
$T$&conserved&conserved\\[5mm]
$C$&conserved&conserved\\[5mm]
$PT$&violated&conserved\\[5mm]
$PC$&violated&conserved\\[5mm]
$TC$&conserved&conserved\\[5mm]
$PCT$&violated&conserved\\\hline
\end{tabular}
\end{center}
concluding that in the 4-dimensional manifold only parity is violated while at each plane for fixed $z$ all the 3-dimensional discrete symmetries
are conserved. These results are not a surprise, external magnetic fields acting on planar systems usually break 4-dimensional parity
and, from the construction of the Lagrangian~(\ref{L_C_phi}), it was expected invariance under the  3-dimensional discrete symmetries.

\section{Conclusions and Outlook}

In this article we have derived a Landau-Ginzburg Chern-Simons model given by a mathematical setup with an internal dynamical pseudo-photon field
which preserve the planar (3-dimensional) discrete symmetries $P$ and $T$, both at the level of the action and
at the level of the electromagnetic equations. Some of the quantitative results obtained are in agreement with the standard fractional
Hall effect description by effective $U(1)$ models, however the physical interpretation in terms of the field contents is distinct, in particular
exist an orthogonal electric field and longitudinal magnetic fields which are usually not present in the standard Hall effect.
Nevertheless the model solves the inconsistence of the
vector/pseudo-vector nature of the electromagnetic equations, in particular the vectorial electric Hall current
is given by a vectorial expression~(\ref{Je_i_Jg_i}) as opposed to a pseudo-vectorial expression~(\ref{hall_orig}),
the model also accounts for the effective charge of anyons being always $e^*=e/(2m-1)$ for any fractional Hall conductance
$\sigma_H=(e/\Phi_0)\,p/(2m-1)$ independently of the value $p$ as has been experimentally verified~\cite{exp_charge}.
These features are mainly due to the existence of both electric and magnetic vortex configurations corresponding, respectively
to the dynamical pseudo-photon and the non-dynamical internal photon. The pseudo-photon electric vortexes
create a statistical electric field orthogonal to the system which may be relevant when computing inter-layer correlations functions
in bi-layer Hall systems having a Bose Einstein Condensate phases~\cite{BEC}. As a preliminary justification for this statement
we recall that from a field theory perspective the interactions are due to the gauge fields, for the particular model discussed
here the interlayer interactions are due to the pseudo-photon gauge field. Also we note that standard 3-dimensional $U(1)$ Maxwell
Chern-Simons theory does not allow to describe orthogonal electric fields, this construction is only achievable by considering extended
gauge theories descriptions of electromagnetism, as is the case of pseudo-photon theory. 
Also we have shown that for the model developed here the Dirac quantization condition for the coupling constants $e$ and $g$~\cite{Dirac,YW}
is equivalent to the quantization of magnetic flux~\cite{London}, here explicit in the quantization of the Hall conductance $\sigma_H$.

We have further analysed thick (4-dimensional) model configurations concluding that for several distinct fine-tune of
the planar 3-dimensional statistical charges the physical charges, as perceived in the 4-dimensional manifold (hence as
measurable outside of the system), are the same. This result allows for distinct interpretations of the possible
planar configurations along the orthogonal direction to the system not giving an unique answer with respect to the possible
microscopical mechanisms within the system. In particular we have obtained either spatial separation of electric and magnetic charges
in the planar system such that no microscopical anyons exist (electric and magnetic excitations are present
in the system at distinct boundaries) either spatial coexistence of both electric and magnetic charges simultaneously
with a Meissner effect such that the external magnetic field does not penetrate the system.
Also our results indicate that, when including gauge fields, a planar description is incomplete and for a full description
of the system (both macroscopically and microscopically) it is required to consider thick models.

Also we note that such distinct characteristics may correspond to distinct phases of the same system such that magnetic vortexes
could be present either confined in bounded pairs such that locally magnetic charge is conserved~\cite{Abrikosov,NO,Nambu},
hence global magnetic charge is null (see also~\cite{BCK} for topological mechanism)
or unconfined. We remark that recently macroscopical magnetic monopole configurations have been experimentally observed~\cite{nature}
in Spin Ice systems, hence, similarly, the model configurations with non-null statistical magnetic charge discussed here may represent
a field configuration that mimics a magnetic monopole through a gauge field configuration. To properly address phase transitions it
is required to include temperature effects through the inclusion of fermionic degrees of freedom. We hope to address this issue somewhere else.

As a final remark we note that in the model discussed in this manuscript the internal photon is considered to be screened. At model level this
was achieved by considering the effective coupling constant $e\alpha$ accounting for the number of unit vortex created by
the internal photon. A possible description of this screening mechanism can be given in a functional formalism
by noting that the vortex solutions correspond to the minimum of the functional energy~\cite{mpl}
in analogy with the Laughlin wave function~(\ref{lgwv_m}) and the internal photon degrees of freedom can be
integrated from the partition function such that the system is consistently described only by the external photon field
(the external electromagnetic fields) and the internal pseudo-photon field. Although in~\cite{mpl} it is only given
an indirect and incomplete prove, the results there indicate a possible explanation for
the effective exclusion of the internal photon from the macroscopical path integral.
Another relevant property of Hall systems not explained by these constructions is the fractional
spin-statistics relation for the excitations in the system, here it is imposed externally by fine-tuning the electric and magnetic
flux of anyons. A possible direction to follow is to consider the homology cycles of the underlying manifold~\cite{Semenoff}.
Although working in a topological trivial manifold, the plane, the effect of considering electrons in the system is equivalent to pierce
the underlying manifold creating non-contractible homology cycle (one for each electron in the system~\cite{Dbranes}).

\vspace{5mm}\noindent {\bf Acknowledgments}\\

This work was supported by SFRH/BPD/34566/2007 from FCT-MCTES.\\


\begin{thebibliography}{99}

\bibitem{IHE_exp} K. von Klitzing, G. Dorda and M. Pepper, \textit{New Method for High-Accuracy Determination of the Fine-Structure Constant Based on Quantized Hall Resistance}, Phys. Rev. Lett. {\bf 45} (1982) 494.

\bibitem{Laughlin_1} R. B. Laughlin, \textit{Quantized Hall conductivity in two dimensions}, Phys. Rev. {\bf B23} (1981) 5632.

\bibitem{FQHE_exp} D. C. Tsui, H. L. Stormer and A. C. Gossard, \textit{Two-Dimensional Magnetotransport in the Extreme Quantum Limit}, Phys. Rev. Lett. {\bf 48} (1982) 1559.

\bibitem{Laughlin_2} R. B. Laughlin, \textit{Anomalous quantum Hall effect: An incompressible quantum fluid with fractionally charged excitations}, Phys. Rev. Lett. {\bf 50} (1982) 1395; \textit{Quantized motion of three two-dimensional electrons in a strong magnetic field}, Phys. Rev. {\bf B27}, 3383.

\bibitem{frac_1} F. Wilczek, \textit{Quantum Mechanics of Fractional-Spin Particles}, Phys. Rev. Lett. {\bf 49} (1982) 957.

\bibitem{Haldane} F. D. M. Haldane, \textit{Fractional Quantization of the Hall Effect: A Hierarchy of Incompressible Quantum Fluid States}, Phys. Rev. Lett. {\bf 51} (1983) 605.

\bibitem{Alperin} B. I. Alperin, \textit{Statistics of Quasiparticles and the Hierarchy of Fractional Quantized Hall States}, Phys. Rev. Lett. {\bf 52} (1984) 1583.

\bibitem{frac_4} D. Arovas, J. R. Schrieffer and F. Wilczek, \textit{Fractional Statistics and the Quantum Hall Effect}, Phys. Rev. Lett. {\bf 53} (1984) 722.

\bibitem{frac_5} D. Arovas, J. R. Schrieffer, F. Wilczek and A. Zee, \textit{Statistical Mechanics Of Anyons}, Nucl. Phys. {\bf B251} (1985) 117.

\bibitem{Berry} M. V. Berry, \textit{Quantal Phase Factors Accompanying Adiabatic Changes}, Proc. R. Soc. Lond. {\bf A 392} (1984) 45.

\bibitem{AB} Y. Aharonov and D. Bohm, \textit{Significance of Electromagnetic Potentials in the Quantum Theory}, Phys. Rev. {\bf 115} (1959) 485.

\bibitem{Simon} B. Simon, \textit{Holonomy, the Quantum Adiabatic Theorem, and Berry's Phase}, Phys. Rev. Lett. {\bf 51} (1983) 2167.

\bibitem{spin_1} F. Wilczek, \textit{Magnetic Flux, Angular Momentum, and Statistics}, Phys. Rev. Lett. {\bf 48} (1982) 1144.

\bibitem{footnote1} Actually one must choose which spin to assign to the particles and accordingly obtain a modified spin-statistics relation and
respective phase rotations~\cite{Haldane,CS_2}. However scalars usually represent spin~0 particles and in this work we adopt this standard choice.


\bibitem{SC_1} V.L. Ginzburg and L.D. Landau, \textit{On the Theory of superconductivity}, Zh. Eksp. Teor. Fiz. {\bf 20} (1950) 1064.
\bibitem{SC_2} L. N. Cooper, \textit{Bound Electron Pairs in a Degenerate Fermi Gas}, Phys. Rev. {\bf 104} (1956) 1189.
\bibitem{SC_3} J. Bardeen, L. N. Cooper and J. R. Schrieffer, \textit{Microscopic Theory of Superconductivity}, Phys. Rev. {\bf 106} (1957) 162; \textit{Theory of Superconductivity}, Phys. Rev. {\bf 108} (1957) 1175.

\bibitem{mass_CS} S. Deser, R. Jackiw and S. Templeton, \textit{Topologically Massive Gauge Theories}, Annals Phys. {\bf 140} (1982) 372-411; Annals Phys. {\bf 185} (1988) 406; Annals Phys. {\bf 281} (2000) 409-449.
\bibitem{CS_1} S. M. Girvin and A. H. MacDonald, \textit{Off-Diagonal Long-Range Order, Oblique Confinement, and the Fractional Quantum Hall Effect}, Phys. Rev. Lett. {\bf 58} (1987) 1252.
\bibitem{CS_2} S.C. Zhang, T. H. Hansson and S. Kivelson, \textit{Effective-Field-Theory for the Fractional Quantum Hall Effect}, Phys. Rev. Lett. {\bf 62} (1989) 82.
\bibitem{CS_3} N. Read, \textit{Order Parameter and Ginzburg-Landau Theory for the Fractional Quantum Hall Effect}, Phys. Rev. Lett. {\bf 62} (1989) 86.
\bibitem{Jain} J. K. Jain, \textit{Composite-Fermion Approach for the Fractional Quantum Hall Effect}, Phys. Rev. Lett. {\bf 63} (1989) 199; \textit{Theory of the fractional Hall effect}, Phys. Rev. {\bf B41} (1990) 7653.
\bibitem{spin_4} C. R. Hagen, \textit{Rotational Anomalies Without Anyons}, Phys. Rev. {\bf D31} (1985) 2135-2136.

\bibitem{spin_2} G. Semenoff, \textit{Canonical Quantum Field Theory with Exotic Statistics}, Phys. Rev. Lett. {\bf 61} (1988) 517.

\bibitem{spin_3} T. H. Hansson, M. Roc\v{e}k, I. Zahed and S.C Zhang, \textit{Spin and Statistics in Massive (2+1)-Dimensional QED}, Phys. Lett. {\bf B214} (1988) 475.

\bibitem{Semenoff} M. Bergeron, G. W. Semenoff and R. J. Szabo, \textit{Canonical BF Type Topological Field Theory and Fractional Statistics of Strings}, Nucl. Phys. {\bf 437} (1995) 695-722, \texttt{hep-th/9407020}.

\bibitem{Polyakov} A. M. Polyakov, \textit{Fermi-Bose Transmutations Induced by Gauge Fields}, Mod. Phys. Lett. {\bf A3} (1988) 325.

\bibitem{CFT} G. W. Moore and N. Seiberg, \textit{Taming the Conformal Zoo}, Phys. Lett. {\bf B220} (1989) 422; S. Elitzur, G. W. Moore, A. Schwimmer and N. Seiberg, \textit{Remarks on the Canonical Quantization of the Chern-Simons-Witten Theory}, Nucl. Phys. {\bf B326} (1989) 108; M. Bos and V. P. Nair, \textit{$U(1)$ Chern-Simons Theory and c=1 Conformal Blocks}, Phys. Lett. {\bf B223} (1989) 61. 

\bibitem{Kogan} I. I Kogan, \textit{The Off-Shell Closed String as Topological Open Membranes. Dynamical Transmutation of World Sheet Dimension}, Phys. Lett. {\bf B231} (1989) 377.

\bibitem{BCK} P. Castelo Ferreira, I. I. Kogan and B. Tekin, \textit{Toroidal Compactification in String Theory from Chern-Simons Theory}, Nucl. Phys. {\bf B589} (2000) 167-195, \texttt{hep-th/0004078}.

\bibitem{Horava} P. Ho\v{r}ava, \textit{Open Strings From Three-Dimensions: Chern-Simons-Witten Theory On Orbifolds}, J. Geom. Phys. {\bf 21} (1996) 1-33, \texttt{hep-th/9404101}; \textit{Strings on World Sheet Orbifolds}, Nucl. Phys. {\bf B327} (1989) 461.

\bibitem{PCT} P. Castelo Ferreira and I. I. Kogan, \textit{Open and unoriented strings from topological membrane. I. Prolegomena}, JHEP 0106 (2001) 056, \texttt{hep-th/0012188}; Erratum ibid, submitted for publication.

\bibitem{Dbranes} P. Castelo Ferreira, I. I. Kogan and R. Szabo, \textit{D-branes in topological membranes}, Nucl. Phys. {\bf B676} (2004) 243-310, \texttt{hep-th/0308101}.


\bibitem{K_CS_1} X. G. Wen, Frank Wilczek and A. Zee, \textit{Chiral spin states and superconductivity}, Phys. Rev. {\bf B39} (1989) 11413.
\bibitem{K_CS_2} X. G. Wen and A. Zee, \textit{Quantum Statistics And Superconductivity In Two Spatial Dimensions}, Nucl. Phys. Proc. Suppl. {\bf 15} (1990) 135.
\bibitem{K_CS_3} B. Blok and X. G. Wen, \textit{Effective Theories of the Fractional Quantum Hall Effect at Generic Filling Fractions}, Phys. Rev. {\bf B42} (1990) 8133; \textit{Effective theories of the fractional quantum Hall effect: Hierarchy construction}, Phys. Rev. {\bf B42} (1990) 8145.
\bibitem{K_CS_4} J. Fr\"ohlich and T. Kerler, \textit{Universality in quantum Hall systems}, Nucl. Phys. {\bf B354} (1991) 369.
\bibitem{K_CS_5} J. Fr\"ohlich and A. Zee, \textit{Large scale physics of the quantum hall fluid}, Nucl. Phys. {\bf B364} (1991) 517.

\bibitem{review} T. Chakraborty and P. Pietil\"ainen, Springer-Verlag 1995; T. Ando, A. B. Fowler and F. Stern, \textit{Electronic properties of two-dimensional systems}, Rev. Mod. Phys. {\bf 54} (1982) 437; Sumathi Rao, \textit{An Anyon Primer}, \texttt{hep-th/9209066}; X.-G. Wen, \textit{Topological orders and Edge excitations in FQH states}, Advances in Physics {\bf 44} (1995) 405, \texttt{cond-mat/9506066}; G. Murthy and R. Shankar, \textit{Hamiltonian theories for the fractional Hall effect}, Rev. Mod. Phys. {\bf 75} (2003) 1101; P. A. Horv\'athy, \textit{Lectures on (abelian) Chern-Simons vortices}, Lectures at NIKHEF (July 2006), \texttt{arXiv:0704.3220v1}.

\bibitem{Abrikosov} A. A. Abrikosov, \textit{On the Magnetic Properties of Superconductors of the Second Group}, Sov. Phys. JETP {\bf 5} (1957) 1174-1182; Zh. Eksp. Teor. Fiz. {\bf 32} (1957) 1442-1452.
\bibitem{NO} H. B. Nielsen and P. Olesen, \textit{Vortex-line models for dual strings}, Nucl. Phys. {\bf B61} (1973) 45.
\bibitem{Nambu} Y. Nambu, \textit{String, Monopoles, and Gauge Fields}, Phys. Rev. D10 (1974) 4262.


\bibitem{Laughlin_3} R. B. Laughlin, \textit{Superconducting Ground State of Noninteracting Particles Obeying Fractional Statistics}, Phys. Rev. Lett. {\bf 60} (1988) 2677.

\bibitem{footnote2} Corresponding in the microscopical quantum theory to one quanta of magnetic flux.

\bibitem{exp_charge} R. G. Clark, J. R. Mallett, S. R. Haynes, J. J. Harris and C. T. Foxon, \textit{Experimental determination of fractional charge e/q for quasiparticle excitations in the fractional quantum Hall effect}, Phys. Rev. Lett. {\bf 60} (1988), 1747; V. J. Goldman, \textit{Resonant tunneling in the quantum Hall regime: measurement of fractional charge}, Surface Science {\bf 361/362} (1996) 1-6; V.J. Goldman, I. Karakurt, J. Liu and A. Zaslavski, \textit{Invariance of charge of Laughlin quasiparticles}, Phys. Rev. {B 64} (2001) 085319.

\bibitem{Sing_2} A. Kato and D. Singleton, \textit{Gauging Dual Symmetry}, Int. J. Phys. {\bf 41} (2002) 1563, \texttt{hep-th/0106277}.

\bibitem{CF} N. Cabibbo and E. Ferrari, \textit{Quantum electrodynamics with Dirac monopoles}, Il Nuovo Cimento {\bf XXIII} No 6 (1962) 1147.

\bibitem{CCN} P. C. R. Cardoso de Mello, S. Carneiro e M. C. Nemes, \textit{Action Principle for the Classical Dual Electrodynamics}, Phys. Lett. {\bf B384} 197, \texttt{hep-th/9609218}.

\bibitem{Sing_1} D. Singleton, \textit{Electromagnetism with Magnetic Charge and Two Photons}, Am. J. Phys. {\bf 64} (1996) 452; \textit{Topological Electric Charge}, Int. J. Theor. Phys. {\bf 34} (1995) 2453, \texttt{hep-th/9701040}.

\bibitem{action} P. Castelo Ferreira, \textit{Explicit Actions for Electromagnetism with Two Gauge Fields with Only one Electric and one Magnetic Physical Fields}, J. Math. Phys. {\bf 47} (2006) 072902, \texttt{hep-th/0510063}.

\bibitem{pseudo} P. Castelo Ferreira, \textit{A Pseudo-Photon in Non-Trivial Background Fields}, Phys. Lett. {\bf B651} 74-78, \texttt{hep-ph/0609239}.

\bibitem{planar} P. Castelo Ferreira, \textit{$U_e(1)\times U_g(1)$ actions in 2+1-dimensions: Full vectorial electric and magnetic fields}, Europhys. Lett. {\bf 79} (2007) 20004, \texttt{hep-th/0703193}.


\bibitem{Schwarz} A. S. Schwarz, \textit{The Partition Function Of A Degenerate Functional}, Commun. Math. Phys. {\bf 67} (1979) 1-16.

\bibitem{Witten} E. Witten, \textit{$Sl(2,Z)$ Action On Three-Dimensional Conformal Field Theories With Abelian Symmetry}, In *Shifman, M. (ed.) et al.: From fields to strings, vol. 2* 1173, \texttt{hep-th/0307041}.

\bibitem{mpl} P. Castelo Ferreira, \textit{Canonical Functional Quantization of Pseudo-Photons in Planar Systems}, to appear in the proceedings of XVI International Fall Workshop on Geometry and Physics, \texttt{http://www.math.ist.utl.pt/~xvi-iwgp/talks/PedroCasteloF.pdf}, \texttt{arXiv:0712.0710};
work in progress.

\bibitem{Dirac} P. A. M. Dirac, \textit{Quantized Singularities in the Electromagnetic Field}, Proc. Roy. Soc. {\bf A113} (1931) 60;  \textit{The Theory of Magnetic Poles}, Phys. Rev. {\bf 74} (1948) 817.

\bibitem{YW} T. T. Wu and C. N. Yang, \textit{Concept of Nonintegrable Phase Factors and Global Formulation of Gauge Fields}, Phys. Rev. {\bf D12} (1975) 3845; \textit{Dirac's Monopole Without Strings: Classical Lagrangian Theory}, Phys. Rev. {\bf D14} (1976) 437

\bibitem{London} W. Meissner and R. Ochsenfeld, \textit{Ein neuer Effect bei Eintritt der Supraleitf�higkeit}, Naturwiss {\bf 21} (1933) 787; H. London and F. London, \textit{The Electromagnetic Equations of the Supraconductor}, Proc. Roy. Soc. (London) {\bf A149} (1935) 71.

\bibitem{footnote3} This issue can be altogether ignored by considering an orbifold under the map~(\ref{PzO}) such that the system is defined in the interval
$z\in[0,\delta_\perp/2]$ and setting either of the boundary conditions expressed in~(\ref{boundary_charges}) for the boundaries at $z=\delta_\perp/2$
and the conditions expressed in equation~(\ref{orb_bc}) as boundary conditions for the orbifold plane (the boundary at $z=0$) being obtained exactly the
same solutions.

\bibitem{Jackson} J. D. Jackson, \textit{Classical Electrodynamics}, Wiley, Third Edition, ,see section 6.10.

\bibitem{BEC} A. H. MacDonald and E. H. Rezayi, \textit{Fractional quantum Hall effect in a two-dimensional electron-hole fluid}, Phys. Rev. {\bf B42} (1990) 3224; J. P. Eisenstein, L. N. Pfeiffer and K. W. West, \textit{Independently contacted two-dimensional electron systems in double quantum wells}, Appl. Phys. Lett. {\bf 57} (1990) 2324; J. P. Eisenstein, G. S. Boebinger, L. N. Pfeiffer, K. W. West and Song He, \textit{New fractional quantum Hall state in double-layer two-dimensional electron systems}, Phys. Rev. Lett. {68} (1992) 1383; X.-G. Wen and A. Zee, \textit{Neutral superfluid modes and magnetic monopoles in multilayered quantum Hall systems}, Phys. Rev. Lett. {\bf 69} (1992) 1811; I. B. Spielman, J. P. Eisenstein, L. N. Pfeiffer and K. W. West, \textit{Resonantly Enhanced Tunneling in a Double Layer Quantum Hall Ferromagnet}, Phys. Rev. Lett. {\bf 84} (2000) 5808; M. Kellog, I. B. Spielman, J. P. Eisenstein, L. N. Pfeiffer and K. W. West, \textit{Observation of Quantized Hall Drag in a Strongly Correlated Bilayer Electron System}, Phys. Rev. Lett. {\bf 88} (2002) 126804; M. Kellog, J. P. Eisenstein, L. N. Pfeiffer and K. W. West, \textit{Vanishing Hall Resistance at High Magnetic Field in a Double-Layer Two-Dimensional Electron System}, Phys. Rev. Lett. {\bf 93} (2004) 036801; E. Tutuc, M. Shayegan and D. A. Huse, \textit{Counterflow Measurements in Strongly Correlated GaAs Hole Bilayers: Evidence for Electron-Hole Pairing}, Phys. Rev. Lett. {\bf 93} 036802; J. P. Eisenstein and A.H. MacDonald, \textit{Bose-Einstein condensation of excitons in bilayer electron systems}, Nature {\bf 432} (2004) 691.

\bibitem{AG} A. P. Balachandran and P. Teotonio-Sobrinho, \textit{The Edge States of the BF System and the London Equations}, Int. J. Mod. Phys. {\bf A8} (1993) 723, \texttt{hep-th/9205116}; M. C. Diamantini, P. Sodano and C. A. Trugenberg, \textit{Gauge Theories Of Josephson Junction Arrays}, Nucl. Phys. {\bf B474} (1996) 641-677, \texttt{hep-th/9511168}; \textit{Superconductors With Topological Order}, Eur. Phys. J. {\bf B53} (2006) 19, \texttt{hep-th/0511192}; \textit{Superconducting Coset Topological Fluids In Josephson Junction Arrays}, J. Phys. {\bf A39} (2006) L253-258; \texttt{hep-th/0703140}.

\bibitem{Aconf} C. Chatterjee and A. Lahiri, \textit{Monopole confinement by flux tube}, \texttt{hep-ph/0605107}; \textit{Confinement of monopole using flux string}, \texttt{arXiv:0705.2635}.

\bibitem{nature} C. Castelnovo, R. Moessner and S. L. Sondhi, \textit{Magnetic Monopoles in Spin Ice}, Nature {\bf 451} (2008) 42; S. T. Bramwell, S. R. Giblin, S. Calder, R. Aldus, D. Prabhakaran and T. Fennell, \textit{Measurement of the Charge and Current of Magnetic Monopoles in Spin Ice}, Nature {\bf 461} (2009) 956.

\end{thebibliography}
\end{document}